\documentclass[11pt]{article}

\usepackage[utf8]{inputenc}    
\usepackage[top=0.75in, bottom=0.75in, left=1in, right=1in]{geometry}
\usepackage{setspace}          
\usepackage{amsmath, amssymb}  
\usepackage{graphicx}          
\usepackage{booktabs}          
\usepackage{caption}           
\usepackage{float}             
\usepackage{natbib}            
\usepackage{url}               
\usepackage{hyperref}          
\usepackage{xr-hyper}          
\usepackage{tabularx}          
\usepackage{titlesec}          
\usepackage{authblk}           
\usepackage{fancyhdr}          
\usepackage{enumitem}          
\usepackage{dcolumn}    
\usepackage{siunitx}    
\usepackage{subfig}
\usepackage{tikz}
\usetikzlibrary{positioning, fit, arrows.meta, calc, backgrounds}
\usepackage{threeparttable}

\providecommand{\Description}[1]{}

\titleformat{\section}{\large\bfseries}{\thesection}{1em}{}
\titleformat{\subsection}{\normalsize\bfseries}{\thesubsection}{1em}{}
\titlespacing*{\section}{0pt}{1.0ex plus 0.3ex minus 0.2ex}{0.5ex plus 0.1ex}
\titlespacing*{\subsection}{0pt}{0.8ex plus 0.2ex minus 0.1ex}{0.4ex plus 0.1ex}
\titlespacing*{\paragraph}{0pt}{0.6ex plus 0.2ex minus 0.1ex}{0.4em}

\newif\ifanon
\anonfalse

\ifanon
\hypersetup{pdfauthor={},pdftitle={},pdfsubject={},pdfkeywords={},pdfcreator={}}
\fi

\title{\textbf{Engagement vs. Commitment: The Economic Trade-Offs of Polarizing News Content}}

\ifanon
\author{}
\else
\author[1]{Shunyao Yan}
\author[2]{Klaus M. Miller\thanks{Yan: syan2@scu.edu; Miller: millerk@hec.fr. Thanks for comments to Peter Ebbes, Brad Shapiro, Puneet Manchanda, Dokyun Lee, Ananya Sen, Michelangelo Rossi, Vincent Lefrere,
Desmond Lo, Christian Peukert, Tobias Kircher,  and participants at the Digital Economics Workshops (Athens 2026; Berlin 2025; UK 2025), EMAC 2026, Marketing Science 2025--2026, Marketing Dynamics 2026, INSEAD-HEC-ESSEC Seminar 2026, Next Gen. of Antitrust Scholars (USC Gould) 2026, Boston University Digital Business Seminar 2025, Workshop on the Economics of Advertising and Marketing 2025, POMS 2025, TPM 2024, Winter AMA 2023, and seminars at emlyon Business School, IIM Bangalore, HEC Paris, Tokyo Chuo University, Weizenbaum Institute Berlin, Goethe University, Santa Clara University, and the University of San Francisco.}}

\affil[1]{Santa Clara University}
\affil[2]{HEC Paris}
\fi

\date{}

\begin{document}

\begingroup
\singlespacing
\maketitle
\endgroup

\vspace{-2em}
\begingroup
\singlespacing
\begin{abstract}
\noindent Content that drives engagement need not be the same content that drives willingness to pay. We study how polarizing content affects engagement (time on site) and commitment (subscriptions and retention) on a major news platform. We measure article-level polarization with deep-learning classifiers and large language models tailored to a multiparty system, and identify causal effects with two complementary instrumental variables: a Bartik instrument exploiting supply-side editorial variation, and an election instrument exploiting demand-side political salience. We find that supply-driven increases in polarizing content raise engagement but not subscriptions. During the high-salience election window, the same content reduces subscriptions and accelerates churn, with affective polarization driving the sharpest divergence. On the mechanism, we find evidence inconsistent with confirmation bias: three pre-determined ideology proxies do not moderate the engagement or subscription effects. By contrast, on ideological dimensions where the publisher covers both sides, exogenous shifts in the publisher's supply of content opposite readers' baseline ideology raise their consumption of that content, consistent with balanced consumption. These results document an asymmetric engagement-commitment trade-off for digital publishers: polarizing content reliably captures attention but does not convert to subscriptions, and actively damages commitment when political salience is elevated. 
\end{abstract}
\endgroup

\vspace{-1em}
{\singlespacing
\par}

\newpage

\section{Introduction}\label{introduction}

News publishers earn revenue from two distinct sources, advertising on each page view and subscription on each retained reader \citep{AndersonCoate2005, Armstrong2006, Wilbur2008, ChaeHaSchweidel2023}. As digital advertising revenue has declined, subscription revenue has surpassed advertising at the median news organization \citep{newman2025dnr}. Conversion from casual reader to paying subscriber is rare, and the lifetime value of a converted subscriber is large relative to the marginal advertising revenue from short-run engagement gains. Once acquired, news subscribers are unusually durable, with annual-plan retention at 71\% for the median news brand and 83\% at top performers \citep{piechota2020retention}. Whether content that raises engagement also raises subscriptions and retention is therefore a first-order question for editorial strategy.

\textit{Polarizing content} is central to this tension. Broadly, we define polarizing content as content that conveys a stance corresponding to one side of a political spectrum or that employs emotionally charged us--versus--them language (we formalize this definition in Section~3). On the engagement side, the case is well established: polarizing language reliably attracts attention online \citep{RathjeVanBavelVanderLinden2021PNAS, BradyEtAl2017PNAS, banerjee2025language, BergerMilkman2012}, and engagement is closely linked to advertising revenue \citep{gentzkow2014trading, Wilbur2008}; whether it also lifts subscriptions and retention, the margins on which publisher revenue now centrally depend, is an open question. If engagement and commitment diverge, the standard editorial heuristic to follow revealed engagement preferences is incomplete.

Why might engagement and commitment diverge for polarizing news content? Prior literature suggests two distinct, and potentially competing, forces shaping consumers’ demand for news. One stream, rooted in confirmation bias theory, finds that consumers are drawn to ``slanted'' media outlets that align with and reinforce their existing political beliefs, suggesting that consumers respond positively to polarizing content, provided that it validates their worldview \citep{gentzkow2010drives, MullainathanShleifer2005}. Consumers’ attraction to such content has well-documented downstream effects on political, economic, and social behavior \citep{DellaVignaFerrara2015, MartinYurukoglu2017,simonov2022frontiers}. Yet a second stream complicates this picture by suggesting that consumers also have strong preferences for balanced and quality information. Empirical evidence shows that the average US consumer does not limit news consumption to a restricted set of ideologically similar outlets and, if anything, favors a balanced mix of sources \citep{GentzkowShapiro2011,Guess2021}. Even in contexts where ideological supply is heavily skewed, as in Russia, consumers' revealed preferences appear driven more by non-ideological factors such as non-political content and third-party referrals than by ideological affinity \citep{simonov2022demand}.

These opposing tendencies, attraction to polarizing content alongside preferences for balance, motivate us to gauge commitment beyond engagement. To do so, we partner with a major European newspaper of record and track each reader simultaneously across engagement, subscription, and retention. Our panel covers every registered user over 40 consecutive weeks, including the focal country's regularly scheduled federal election, and records each article a user reads together with their subscription and retention outcomes. Holding the publisher's pricing and content delivery fixed throughout, exogenous variation in polarizing supply and demand identifies its causal effect on both engagement and commitment within the same user, isolating a within-publisher content margin that cross-platform and cross-outlet designs cannot reach. Our central finding is an asymmetric engagement-commitment trade-off: polarizing content lifts time on site but fails to convert to subscriptions, and actively damages commitment when political salience is elevated. 

In conducting our analysis, we overcome two main empirical challenges. The first lies in measuring polarization in news content. Existing text-based measures collapse polarization onto a single left--right label per article. Two features of our setting make that approach restrictive. First, polarization is multi-faceted. Affective polarization, the widening of in-group versus out-group affect, travels through emotionally charged language and has become a defining feature of contemporary polarization \citep{IyengarSoodLelkes2012,iyengar2019origins,boxell2024cross}, yet it is largely missed by lexical and stance metrics built around issue positions \citep{gentzkow2010drives,GentzkowShapiroTaddy2019}. Second, news articles can quote, paraphrase, or refute partisan language without endorsing it, while factual policy coverage can take a clear issue position without sounding like partisan discourse. A single article-level label therefore conflates two distinct signals: how closely the language aligns with party rhetoric, and what stance a reader would infer from the article as a whole. This distinction matters: engagement may respond to rhetorical cues and us--versus--them language, while subscription and retention may depend on whether readers perceive the article and publisher as balanced and credible. 

To address these measurement challenges, we produce two complementary article-level measures of polarization: a \textit{party-referenced} measure of what is inherently polarizing in the language, based on a hierarchical BERT classifier (Bidirectional Encoder Representations from Transformers; a transformer-based deep-learning model) trained on parliamentary speech, and a \textit{reader-referenced} measure of the polarization a reader would perceive, from a large language model (LLM) scoring each article along affective and three ideological dimensions. Section~\ref{two-component-framework-measuring-polarization} details the framework. Pairing both reference systems on the same articles, which no prior measure does, lets us identify which dimension drives the engagement-commitment trade-off.

The second challenge lies in identification: Isolating the causal effect of consuming polarizing content from selection bias is a well-known challenge in media effects research \citep{DellaVignaKaplan2007,arceneaux2013changing}. Randomized experiments, while considered the gold standard for causal inference, face external validity concerns in this domain: for example, \citet{broockman2025consuming} show that participants tend to revert to their preferred media sources after an experiment, making it unclear whether experimental results generalize to real-world consumption.

Our solution is a quasi-experimental instrumental-variable (IV) strategy that exploits plausibly exogenous variation from both the supply and the demand side. We construct two theoretically distinct instruments. First, a Bartik-style instrument (the Bartik instrument) interacts users' stable topic preferences with weekly shifts in the publisher's supply of polarizing content, driven by editorial planning and exogenous news events. Second, an election-driven demand shock (the Election instrument) raises political salience only for users in the affected country; users in a linguistically similar neighbor country, an institutional feature common in multilingual European states, serve as a comparison group.

Our findings document an asymmetric trade-off: exogenous increases in polarizing content raise time on site but lower the probability of subscribing and increase the probability of churn, with the commitment penalty sharpest during the political salience of the federal election. Affective polarization drives the largest divergence: the Bartik instrument for affectively polarizing content draws substantial additional engagement (semi-elasticity $\approx 11.6\%$), but during the election window, an extra click on such content reduces the weekly subscription conversion probability by about 1.6 percentage points, more than two and a half times the 0.63\% weekly base rate. For ideological content, supply-driven engagement responses are narrower and do not translate into measurable subscription gains. Using three pre-determined ideology proxies (residential location matched to local vote shares as an exogenous geographic proxy, and pre-period reading classified independently by our LLM scorer and BERT classifier as two revealed-preference proxies), we find that the engagement-commitment trade-off does not vary with these ideology proxies, nor does polarizing content push readers deeper into the subscription funnel. What readers do respond to is exogenous variation in the publisher's supply of content opposite their pre-period lean on three ideological dimensions where the publisher covers both sides (Economic, Globalization, and Environment), evidence consistent with balanced consumption rather than confirmation bias. 

This study makes three contributions. First, we extend prior work on polarizing content beyond attention and political behavior to the subscription and retention margins that determine subscriber lifetime value at digital publishers. We calculate that the implied subscription loss from one additional polarizing click exceeds marginal advertising revenue by 59 to 176 times even at the conservative end of the confidence interval (Section~6 details), overturning the editorial premise that engagement metrics proxy publisher value. Second, we adjudicate the demand-side mechanism in favor of balanced consumption over confirmation bias: heterogeneity by reader ideology shows no systematic moderation effects, and exogenous counter-side supply shocks raise counter-side reading at least one-for-one. Third, we introduce a validated article-level polarization framework that decomposes polarization into one affective and three ideological dimensions, combining a party-referenced BERT classifier with a reader-referenced LLM scorer, which, to our knowledge, is the first such framework and makes the affective-versus-ideological asymmetry behind the commitment penalty operationally testable. 

\section{Related Literature}\label{sec:related-literature}

Our study draws on and contributes to three main streams of literature.

First, we expand the literature on the effects of polarizing content and partisan media. A long-standing body of causal evidence establishes that exposure to partisan-media content reshapes political behavior and economic and social outcomes, from vote shares to investment decisions and public-health compliance \citep{DellaVignaFerrara2015, simonov2022frontiers, DellaVignaKaplan2007,MartinYurukoglu2017}. On digital platforms, polarizing content is unusually effective at capturing attention. Out-group animosity, moral-emotional language, and high-arousal affect all systematically increase social-media engagement, diffusion, and click-through \citep{RathjeVanBavelVanderLinden2021PNAS, BradyEtAl2017PNAS,BergerMilkman2012}. \citet{braghieri2025article} further document that within-outlet variation in article-level slant accounts for the majority of the polarization observed in news consumption across major U.S. outlets, establishing that the politically consequential variation in news content operates within, rather than across, publishers. Taken together, this literature establishes that polarizing content reliably captures attention, diffuses widely, and shifts downstream political attitudes, yet it does not examine how such content causally affects the revenue margins that sustain publishers. Engagement is informative about advertising revenue only through the assumed pricing relationship between user attention and ad value \citep{gentzkow2014trading, Wilbur2008}, not from a direct measurement of ad revenue itself. Subscription and retention, the margins that now account for the majority of publisher revenue, have not yet been examined in the context of polarization.

Second, our work contributes to the literature on the factors driving demand for digital media. Existing work has identified consumer demographics \citep{RustAlpert1984}, content psychological themes \citep{ToubiaIyengarBunnellLemaire2019}, emotional intensity \citep{BergerMilkman2012}, prior content choices \citep{WoolleySharif2022}, ad blockers \citep{yan2022does, todri2022impact}, headline language \citep{banerjee2025language}, and smartphone access \citep{aridoretal2025}. The closest precedent is \citet{simonov2022demand}, who score articles along a pro-/anti-government dimension and use individual-level browsing data to decompose Russian news demand into ideological and non-ideological drivers, identifying the latter as dominant.

We extend this line of research in two ways. First, we add affective polarization, the widening of in-group versus out-group affect, as a distinct article-level dimension, complementing the ideological dimensions that prior work has used. Our measurement validation also shows that the affective dimension is genuinely separable from both ideological direction and from generic emotional tone. Second, we observe the same users on the subscription and retention margins, not only clicks or outlet choice, so we can ask whether the content attribute that raises the click margin also raises the commitment margin within a single publisher's editorial supply. By measuring both margins jointly at the article level, we identify a content attribute whose contribution to engagement diverges from its contribution to commitment. 

Beyond these demand drivers, a growing literature studies the economics of digital news subscriptions directly. \citet{ChiouTucker2013} use a local-paper paywall rollout to document a sharp drop in visits and disproportionate exit by young readers; \citet{PattabhiramaiahSriramManchanda2019} estimate that the \textit{New York Times} paywall depressed content demand but lifted subscriptions; \citet{AralDhillon2021} extend this with a quasi-experiment varying paywall quantity and exclusivity; and \citet{ChaeHaSchweidel2023} examine paywall suspensions during major events. These studies vary paywall design while holding content fixed. Our setting reverses the variation: we hold paywall design fixed and shock content composition, isolating the content-margin question and tracking its consequences across engagement, subscription, and churn simultaneously. 

Third, we build on a large interdisciplinary literature that measures political polarization from textual data (see \citealt{PuglisiJr2015} for a review). Earlier approaches relied on differential use of partisan phrases \citep{GentzkowShapiroTaddy2019}, selective coverage \citep{PuglisiJr2011}, and audience following patterns \citep{SchoenmuellerNetzerStahl2019}, mostly aggregating slant to the outlet level. More recent work applies LLMs to score slant at the article level \citep{yoganarasimhan2024feeds, le2025positioning, braghieri2025article}. \citet{yoganarasimhan2024feeds} introduce a scalable LLM-based polarization score for U.S. news, benchmark LLM article scores against human survey judgments, and establish both their accuracy and scalability. We extend this article-level LLM approach in two ways: we decompose polarization into separate affective and ideological-issue components rather than a single left-right score, and we pair the LLM scorer with a supervised BERT classifier trained on parliamentary speech, so each article is described in two complementary reference systems. To our knowledge, no prior work jointly produces these two measures at scale to estimate their causal effects on user behavior, nor measures polarization in a European multiparty setting.

\section{Empirical Setting, Datasets, and Variable Construction}\label{empirical-setting-dataset-and-variable-construction}

\subsection{Publisher and Institutional Setting}\label{publisher-setting}

This study examines how exposure to polarizing articles affects user behavior on a major European news platform. The publisher is headquartered in a Western European country, where the majority of its users are located. We observe user behavior over the course of our 40-week observation window, during which a regularly scheduled federal election was held in the focal country (but not in a neighboring, same-language country in which a substantial minority of the publisher's user base is located), providing plausibly exogenous variation in political salience.

Our publisher has long been regarded as the only national ``newspaper of record'' in its linguistic area in the focal country. Its reputation is comparable to that of \emph{The New York Times}, the \emph{Financial Times}, or \emph{The Guardian}. In addition to a printed newspaper, our publisher runs a digital platform that publishes daily news across topics including politics and economics. The news platform ranks among the top 10 in its country in weekly usage. At the time of our study, around 80\% of the traffic to the platform came from its own country. The publisher markets itself to same-language neighboring-country readers as offering an independent, outside perspective on their domestic affairs, positioning its coverage as a complement to, rather than a substitute for, local media in the neighboring country. This editorial positioning means that neighboring-country users primarily value the publisher for its international and analytical content rather than its focal-country political coverage.

Our observation window spans 40 consecutive weeks in the late 2010s.\footnote{The exact dates are withheld under the publisher non-disclosure agreement (NDA).} Unlike major US outlets, where consumers sort across publishers along partisan lines \citep{GentzkowShapiro2011}, the publisher's long-standing role as a leading national newspaper of record attracts a broad, ideologically diverse readership, mitigating concerns about ideological selection into the news platform. This diverse readership, combined with the publisher's newsroom practice of covering positions across the political spectrum on major policy issues, provides both the cross-user heterogeneity in political priors and the supply variation required for our demand-side mechanism tests of confirmation bias and balanced consumption (Section~\ref{mechanism}).

The publisher we study here is a legacy ``agenda-setting'' outlet with a newsroom culture that emphasizes policy depth over click-through optimization. Week-to-week variation in the topic mix is driven primarily by the external news environment that editors react to under fixed weekly planning routines. According to internal guidelines and interviews with the editors, topical emphases are set in weekly editorial conferences on the basis of newsworthiness and forward-planned series, not real-time page-view metrics; consistent with this, our shock-balance and reverse-causality tests in Section~\ref{instrument-validity} find no detectable response of supply to lagged demand. Major news events, most notably the federal election, are mapped out months in advance through dedicated series and pre-written explainers, so that the timing and intensity of election-related coverage are set by the political calendar rather than by readers' contemporaneous interest. Because breaking news remains a small share of total output and the publisher did not employ algorithmic headline testing or dynamic paywall adjustments during our observation window, the week-to-week variation in polarizing supply reflects editorial agenda-setting and exogenous news shocks rather than contemporaneous user demand. 

The platform's content delivery is almost entirely editorially curated: the main homepage, section pages, and in-article ``Related stories'' links are all driven by editorial topic tagging. The sole exception is a single personalized homepage box near the bottom of the homepage, which surfaced articles based on individual reading history. The platform did not employ collaborative filtering or user-level content targeting beyond this single module, and the publisher does not geo-target content: aside from that specific module, all users, regardless of country, see the same editorially curated homepage, section pages, and article recommendations. This combination of baseline editorial curation and uniform content delivery limits the scope for algorithmic amplification of polarizing consumption and ensures that domestic and foreign users face an identical base content catalogue so that any differential consumption during the election reflects demand-side differences rather than supply-side curation.

The news platform generates revenue from two main sources: display advertising, averaging about five ad slots per page, and paywall subscriptions, priced at approximately~\$35 per month. The paywall operates on a metered model, allowing unregistered users to read up to five free articles or registered users to read up to ten free articles each month before access becomes restricted (meaning the full article view is limited). Users can then purchase an all-access digital subscription to continue reading on the platform. Subscription prices and plan features remained unchanged throughout the 40-week observation window, ruling out pricing variation as a confound. 

Because unregistered users can read up to five articles per month without creating an account, our sample of registered users likely over-represents more engaged and frequent readers relative to the full user population. Registered users have revealed a higher baseline attachment to the platform, making them a conservative sample for detecting content-driven deterrence: any negative effect we estimate is identified on a population that begins with above-average willingness to engage with and pay for the publisher's content.

\subsection{Datasets}\label{datasets}

Our analysis combines three complementary datasets. First, we use a complete archive of 17,776 parliamentary speeches from the focal country during the observation year as training data for our hierarchical BERT classifier of polarizing language. The classifier learns to identify polarizing rhetoric from a balanced sample of Left (5,915) and Right (5,931) speeches, with Center and Executive speeches providing the non-polarizing comparison cases (Online Appendix~OA1 provides corpus details and summary statistics). Second, we use 40 weeks of user clickstream logs and purchase records from our publisher, including article views, session length, purchase events, and contract duration. Third, we compile the publisher's complete archive of articles with timestamps. The resulting flow of articles allows us to characterize week-to-week movements in topic-specific supply and to construct the Bartik instrument by interacting users' stable topic preferences with platform-wide shifts in polarizing supply.

\subsection{Measuring Polarization: Two-Component Framework}\label{two-component-framework-measuring-polarization}

Two features of news content make standard polarization measures restrictive for our setting. First, polarization is multi-faceted: an affective (us--versus--them) component can vary independently of issue positions \citep{iyengar2019origins}, and the ideological space itself is multidimensional (e.g., Economic, Globalization, Environment) in European settings \citep{kriesi2006globalization} and increasingly in the U.S. \citep{carmines2012leftright}, so collapsing both onto a single left--right scale discards real variation. Second, document-level measures (whether dictionary-based, embedding-based, or supervised) typically return one label grounded in a single reference system \citep{gentzkow2010drives}. For analyzing news articles, this can blur two distinct directional signals: the degree to which the language aligns with partisan rhetoric and the stance a knowledgeable reader would attribute to the article as a whole. For downstream analysis, that blur can make it difficult to know whether user responses are tied to party-referenced rhetorical alignment or to the article's reader-referenced stance and its affective tone.

We therefore adopt a multidimensional framework that produces, on the same articles, two complementary measures of polarization grounded in different reference systems. The first, the \textit{party-referenced} measure, assesses left--right alignment based on similarity to parliamentary rhetorical patterns. The second, the \textit{reader-referenced} measure, assesses left--right stance on multiple issues at the article level together with affective polarization (us--versus--them emotional intensity). These two sets of measures are related but not identical: quotation, refutation, and contrastive framing can yield strong party-referenced scores alongside a different reader-referenced stance. The two approaches are complementary by design. The supervised BERT classifier is trained on labeled parliamentary speech with ground-truth stance labels, making it well-suited for detecting rhetorical alignment with partisan discourse, a task for which fine-tuned classifiers remain the standard tool. However, it cannot assess dimensions beyond its training labels (notably affective polarization) and may overweight lexical cues when articles quote or paraphrase partisan language without endorsing it. The LLM compensates for these limitations by applying general reasoning to assess overall article stance and affect, but lacks the grounding in a specific political reference system that supervised training provides. Using both yields two independent measures of polarization whose convergence reinforces validity and whose divergence, common in quote-heavy or contrastively framed pieces, creates informative variation for downstream analysis.

We operationalize the \textit{party-referenced} measure with a hierarchical BERT classifier that uses staged decisions, polarization detection followed by left/right stance classification, with stance supervision and hard-negative pairs (left and right perspectives on the same issue), steering the model to learn argumentative direction while reducing its reliance on the mere co-occurrence of partisan lexicon. To mitigate domain shift from parliamentary speech to journalism, we perform domain-adaptive pretraining on the publisher's news archive from two years prior to the observation window (preserving temporal integrity). Online Appendix~OA2 provides full details on model architecture, training, domain adaptation, optimization, evaluation, and benchmarking.

Following \citet{yoganarasimhan2024feeds}, we operationalize the \textit{reader-referenced} measure with a state-of-the-art LLM that scores each article on a 1--5 scale, where 1 anchors the left-leaning (or low affective intensity) pole and 5 the right-leaning (or high affective intensity) pole. Two features extend their design. First, we score four dimensions on separate 1--5 scales rather than a single left--right axis: Affective intensity (us--versus--them rhetoric), Economic (Left--Right), Globalization (Globalist--Nationalist), and Environment (Pro-environment--Pro-growth). This decomposition accommodates the European multi-party setting and isolates affective intensity from ideological direction. Second, every scale point on every dimension is drawn from canonical political-science operationalizations of the underlying constructs, so the rubric is a transcription of established definitions rather than ad-hoc prompt language. Grounding the model's reasoning path in this literature strengthens cross-run replicability under varied decoding configurations and yields inter-scorer agreement of $\kappa_w = 0.63$--$0.78$ on our corpus (OA5.1). The LLM thus receives a scoring rubric with a structured reasoning template requiring, for each dimension, a numerical score, an evidence-based rationale, and a confidence rating. We set temperature to zero for maximum output stability and reproducibility. After systematic benchmarking of leading LLMs (GPT, Claude, Gemini), we selected Gemini 2.5 Pro based on its superior alignment with human expert annotations. Online Appendix~OA5 provides the complete scoring rubric, prompt text, and model selection details.

\begin{figure}[htbp]
\centering
\scalebox{0.9}{\begin{tikzpicture}[
  node distance=3mm,
  box/.style={align=center, inner sep=3pt, font=\footnotesize},
  hdr/.style={box, font=\bfseries\footnotesize},
  itlabel/.style={font=\footnotesize\itshape},
  arr/.style={-{Latex[length=2mm,width=1.5mm]}, thin}
]
\def\colw{0.38\linewidth}
\def\fullw{0.86\linewidth}
\def\colgap{16mm}

\node[draw, box, text width=0.4\linewidth] (corpus) {%
  \textbf{News article corpus}\\[1pt]
  62{,}801 articles, 40-week window};

\node[itlabel, below=6mm of corpus.south, anchor=east, xshift=-\colgap] (lab_left) {Party-referenced measure};
\node[itlabel, below=6mm of corpus.south, anchor=west, xshift=\colgap] (lab_right) {Reader-referenced measure};

\draw[arr] (corpus.south) -- (lab_left.north);
\draw[arr] (corpus.south) -- (lab_right.north);

\node[hdr, text width=\colw, below=2mm of lab_left] (bert_hdr) {Hierarchical BERT classifier};
\node[box, text width=\colw, below=1mm of bert_hdr] (bert_body) {%
  Trained on parliamentary speech\\[1pt]
  Two-stage classification:\\
  polarization, then stance};

\node[hdr, text width=\colw, below=2mm of bert_body] (bert_out) {Output: Left / Right / Neutral};
\node[itlabel, text width=\colw, below=1mm of bert_out, align=center] (bert_interp) {%
  Categorical measure of\\rhetorical alignment};

\draw[arr] (bert_body.south) -- (bert_out.north);

\node[hdr, text width=\colw, below=2mm of lab_right] (llm_hdr) {LLM-based scoring};
\node[box, text width=\colw, below=1mm of llm_hdr] (llm_body) {%
  Structured evaluation of\\articles across four dimensions};

\node[hdr, text width=\colw, below=2mm of llm_body] (llm_aff) {Affective dimension};
\node[box, text width=\colw, below=1mm of llm_aff] (llm_aff_body) {%
  Us--versus--them intensity (1--5)};

\node[hdr, text width=\colw, below=2mm of llm_aff_body] (llm_ideo) {Ideological dimensions (1--5)};
\node[box, text width=\colw, below=1mm of llm_ideo, align=center] (llm_ideo_body) {%
  Economic:\\
  Left $\leftrightarrow$ Right\\[1pt]
  Globalization:\\
  Globalist $\leftrightarrow$ Nationalist\\[1pt]
  Environment:\\
  Pro-environment $\leftrightarrow$ Pro-growth};

\draw[arr] (llm_body.south) -- (llm_aff.north);

\node[box, text width=\colw, below=1mm of bert_interp] (bert_pad) {\phantom{X}};
\coordinate (llm_bottom) at (llm_ideo_body.south);
\path (bert_pad.north) -- (bert_pad.north |- llm_bottom) coordinate (bert_match_bottom);

\begin{scope}[on background layer]
  \node[draw, fit=(bert_hdr)(bert_body)(bert_out)(bert_interp)(bert_match_bottom), inner sep=4pt] (left_col) {};
  \node[draw, fit=(llm_hdr)(llm_body)(llm_aff)(llm_aff_body)(llm_ideo)(llm_ideo_body), inner sep=4pt] (right_col) {};
\end{scope}

\coordinate (bot_y) at ($(left_col.south)!0.5!(right_col.south)$);
\node[draw, box, text width=\fullw, below=6mm of bot_y, align=center] (vec) {%
  \textbf{Article-level polarization vector}\\[2pt]
  2 categorical indicators (BERT: Left, Right)\\
  1 Affective indicator + 6 ideological indicators (LLM)\\[2pt]
  \itshape Aggregated to user-week level for empirical analysis};

\draw[arr] (left_col.south) -- (left_col.south |- vec.north);
\draw[arr] (right_col.south) -- (right_col.south |- vec.north);

\end{tikzpicture}}
\caption{Multi-dimensional Measurement of Political Polarization}
\label{fig:twodim-polarization}
\end{figure}
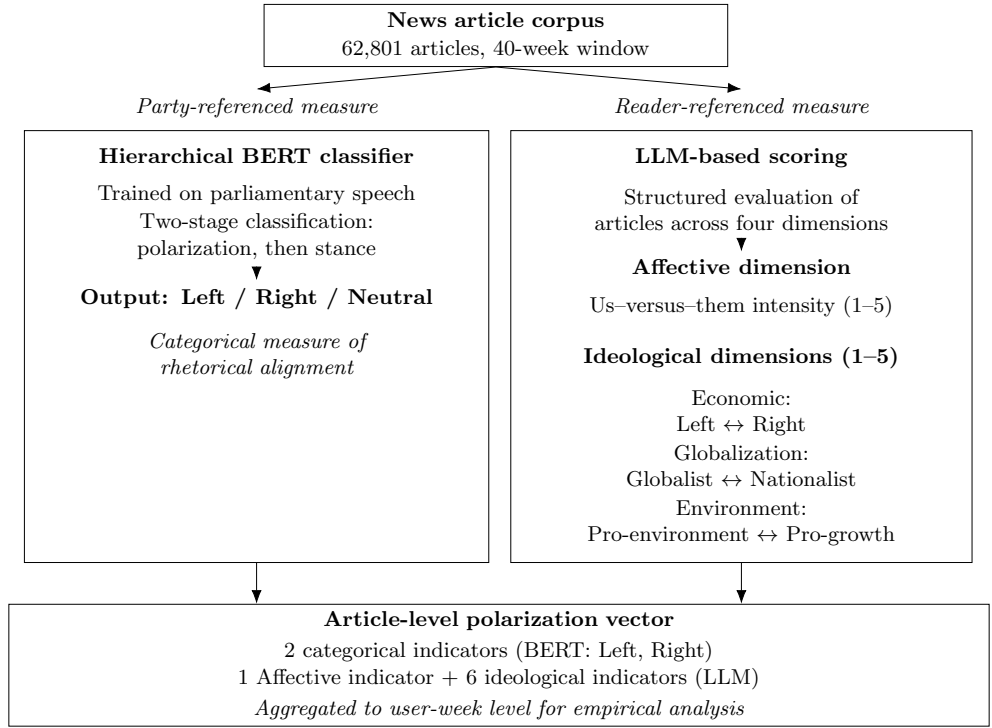

Figure~\ref{fig:twodim-polarization} summarizes the framework: the party-referenced measure (left) produces categorical BERT-Left/BERT-Right/Neutral labels, while the reader-referenced measure (right) scores each article on one affective and three ideological dimensions (Economic, Globalization, Environment), each on a 1--5 scale.

We validate both the BERT and LLM measurements along three dimensions. First, \emph{internal and external validity}: the BERT classifier achieves high in-sample performance and generalizes to out-of-distribution (OOD) articles from ideologically distinct outlets. Second, \emph{reliability}: scores are stable across repeated LLM scoring runs and across three independent LLM families (GPT, Claude, Gemini), with all pairwise expert agreements exceeding 97\%. Third, \emph{discriminant validity}: the affective and ideological dimensions are empirically separable from each other and from general sentiment. Full details and a summary table appear in Online Appendix~OA5, with the main validation results reported below.

The BERT model achieves in-sample $F_1 = 0.922$ (stage~1 binary detection) and macro-$F_1 = 0.938$ (stage~2 directional classification). To assess generalization, a domain expert curated 99 news articles from two outlets with known ideological positions (47 left-leaning, 52 right-leaning) of our focal country. These outlets are not included in the BERT training data. We evaluate the trained BERT classifier on this out-of-distribution (OOD) corpus across five stratified random splits to obtain a standard error on each metric, yielding $F_1 = 0.746$, accuracy $= 75.4\%$ for left-right stance classification, alongside an 88.5\% recall that indicates our BERT classifier rarely misses polarizing content despite domain shift.

Our primary LLM measure, Gemini~2.5~Pro, ranked first on the Chatbot Arena leaderboard (lmarena.ai; over six million human preference votes; \citealt{chiang2024chatbot}) during our analysis period. To verify that scores reflect genuine article properties rather than model-specific biases, we submitted the same 99 OOD articles to two additional LLM systems built on distinct architectures and training pipelines: OpenAI's o3 and Anthropic's Claude~3.7~Sonnet. All pairwise agreements with the human expert exceed 97\%.

\subsection{Application to News Supply}\label{application-news-supply}

We apply our measurement framework to all 62{,}801 articles published during the observation window.

Applying the BERT classifier, we segment articles exceeding the transformer context window into
overlapping chunks and aggregate to the article level using a two-gate
rule that discards low-signal chunks and requires both a posterior
threshold ($\geq 0.55$) and high-confidence chunk consensus
($\geq 75\%$) before assigning a Left or Right label (Online
Appendix~OA2 details the chunking procedure and sensitivity analyses).
Of all articles, 21.7\% are classified as left-leaning and 24.2\% as
right-leaning, with a mean confidence of 0.918; sensitivity analyses
using alternative aggregation rules yield $>95\%$ agreement on
Left/Right assignments.

Applying Gemini~2.5~Pro with our multidimensional rubric to the same
corpus yields the following supply distributions. Only 4.8\% of
articles score as highly affective ($\geq 4$), confirming that strongly
confrontational us--versus--them language is rare in the supply,
consistent with our publisher's positioning as a newspaper of record.
On the Economic dimension, 8.3\% of articles lean left ($\leq 2$)
and 9.6\% lean right ($\geq 4$), consistent with the publisher's economically-liberal editorial profile. Globalization shows a globalist
tilt: 14.4\% globalist ($\leq 2$) versus 7.3\% nationalist
($\geq 4$), consistent with the publisher's substantial international
news coverage. On the Environment dimension, 5.8\% score as
pro-ecology ($\leq 2$) and 2.0\% as pro-growth ($\geq 4$). Online
Appendix~OA3 provides matched examples of articles scoring at
opposite ends of each dimension, and Online
Appendix~OA4 summarizes the full supply distribution. To illustrate, an article reporting on a policy debate using neutral, factual language scores low on the Affective dimension (score~$= 1$), while an article on the same topic framing the debate as a confrontation between ``elites'' and ``ordinary citizens'' scores high (score~$= 5$). On the Economic dimension, an article advocating market deregulation scores right (score~$= 5$), while one calling for expanded social insurance scores left (score~$= 1$).

Two features of the news supply matter for identification. The prevalence of polarizing content varies substantially across topics: Opinion and National Politics carry the highest shares of affectively and ideologically polarizing articles, while Science and Technology carry the lowest, generating the cross-topic supply variation that the Bartik instrument exploits (Online Appendix~OA4 reports the dimension-by-dimension supply distribution; OA6.3 reports the underlying pre-period topic shares). At the same time, the near-symmetric split of BERT-classified articles and the
representation of both ideological poles across all LLM dimensions
confirm that its news supply spans a broad ideological spectrum.
This variation arises naturally from the publisher's role as a
newspaper of record: political reporting covers parties and policies
across the aisle, business coverage includes both pro-market and
regulatory perspectives, and opinion sections feature guest
contributors from diverse viewpoints. 

As our measurement framework decomposes polarization into affective
and ideological dimensions, we demonstrate that these dimensions are
empirically distinct: First, on
the full corpus of 62{,}801~articles, pairwise Spearman correlations
between affective intensity and each ideological dimension are near
zero ($\rho = -0.09$ to $0.05$; OA5.5). Second, Claude's independent left--right
score correlates strongly with Gemini's Economic dimension
($\rho = 0.87$) but only modestly with Gemini's affective dimension
($\rho = 0.25$), confirming from a third independent system that
ideological direction and affective intensity capture distinct article
properties (OA5.4). Third, the inter-scorer
correlation matrix exhibits a multitrait--multimethod pattern
\citep{campbell1959convergent}: same-dimension correlations across
scorers are substantial ($\rho = 0.64$--$0.74$), while all
cross-dimension pairs involving affective and ideological scores fall
below $|\rho| < 0.10$ (OA5.5). Finally, to rule out the possibility that our affective measure simply
proxies for generic emotional valence, we correlate LLM affective
scores with a pre-trained language-specific transformer-based
sentiment classifier for our focal language, applied to article text (512-token sliding
windows, average pooling). The correlation is modest
($\rho = 0.29$, $p < 0.001$), leaving over 91\% of variance
unexplained; meanwhile, sentiment is essentially uncorrelated with all
three ideological dimensions ($|\rho| < 0.04$; OA5.7).

\subsection{Dependent and Independent Variables}\label{dependent-independent-variables}

\noindent\textbf{Clickstream data.} Our dataset, aggregated at the user--week level, comprises pseudonymized clickstream data with unique identifiers for all registered users who visited the news platform during our 40-week observation window (n = 119{,}913); these users account for approximately 20\% of total visits to the platform. Our focus on registered users enables us to track each user individually over time, providing a unique panel setting. We exclude week~1 from formal analysis because the publisher introduced a new tracking system during that week, resulting in a three-day loss of observations and a non-comparable partial week. The resulting panel is unbalanced: a user-week observation exists only when the user visits the platform, so that inactive weeks are not represented. Table~\ref{tab:summary_stats} reports summary statistics for the full panel of 1{,}098{,}857 user-week observations across 119{,}913 users over these 39~weeks (an average of roughly 9.2~observed weeks per user). 

\noindent\textbf{Dependent variables.} We specify three dependent variables to capture the publisher's primary revenue streams.

For advertising revenue, the central metric is user engagement (i.e., attention). While traditional models often use page views as a proxy for ad impressions, modern digital advertising is increasingly sold based on ``time-in-view'' and user engagement depth \citep{goldstein2015time, simonov2025attention}. A single, long-scrolling article page can generate multiple ad impressions through lazy-loading and refresh mechanisms, and video ads are sold on a time basis. Therefore, we select time on site as our primary proxy for ad revenue. It is a more comprehensive measure of the valuable attention that can be monetized. Given its highly skewed distribution, we use the inverse hyperbolic sine transformation (asinh) in our regression models later. This standard technique is robust to zero values and provides coefficients that can be interpreted as semi-elasticities.

For subscription revenue, we focus on conversion into active subscribers. Our subscription dependent variable, \textit{Subscription}, is a binary indicator equal to ~1 if the user placed a new subscription order online in week ~$t$, and ~0 otherwise. To provide a clean measure of subscription, we estimate the subscription models on a hazard sample restricted to user-weeks in which the user has not yet placed a subscription order as of week~$t$. This restriction avoids counting already-converted users in the denominator, ensuring that we estimate the effect on the relevant risk population. In our estimation sample, the weekly subscription base rate is 0.63\%. We later estimate the effect of interest using a Linear Probability Model (LPM) with high-dimensional fixed effects, a robust and common approach for binary panel outcomes. Because subscription decisions may respond to content exposure with a delay, we also estimate a distributed-lag specification with two weekly lags under both IVs. Under both IVs, all nine cumulative three-week subscription effects are null after multiple-testing correction. Lagged exposure does not propagate to the contemporaneous subscription decision (Online Appendix~OA13.4.2).

For churn behavior, we link registered users in the clickstream panel to the publisher's customer relationship management (CRM) subscription records, which report start and end dates of all contracts through the end of our observation window. To avoid classifying administrative or payment-processing noise as churn, we treat short gaps of up to seven days between consecutive subscription intervals as continuous coverage; only gaps longer than seven days are treated as churn from paid subscriptions. Operationally, we construct a weekly churn indicator that equals one in the week containing the first uncovered day after a paid subscription period ends, and zero otherwise. Users whose paid subscription continues through the end of the observation window are treated as right-censored (no churn observed within the sample period). For the churn analysis, we restrict the sample to the hazard set of active subscribers: subscriber-weeks in which the user is actively covered by a paid subscription, so that the churn outcome is defined only for those who could plausibly churn. Crucially, this risk set is constructed directly from the CRM records and spans all calendar weeks of a user's subscription spell, regardless of whether they visited the platform that week. This ensures that ``silent'' churn events, where a user stops visiting prior to cancellation, are fully captured. In our estimation sample, the weekly churn base rate is 0.27\%.

\noindent\textbf{Independent variables.} Our primary independent variables are the absolute counts of ``polarizing'' articles clicked by a user in a given week. To identify an article as ``polarizing'', we assign it a set of nine binary indicators of polarization, corresponding to the scores generated by our two-component framework outlined in the previous section. For the party-referenced measure (BERT classifier), we use two indicators, corresponding to whether the model classified the article as left-leaning or right-leaning. For the reader-referenced measure (LLM scorer), we use seven indicators: one indicator corresponding to the affective polarization dimension (takes a value of 1 if the affective polarization score $\geq 4$); and two indicators for each dimension of ideological polarization, corresponding to the two poles of that dimension (economic left vs.\ economic right: value of 1 if the LLM score is $\leq 2$ or $\geq 4$, respectively; globalist/nationalist: LLM score $\leq 2/\geq 4$; pro--environment/pro--growth LLM score $\leq 2/\geq 4$). We consider each of these indicators separately in our analysis because heterogeneity across the affective and three ideological axes is itself central to our findings, the dimensions are empirically distinct constructs (OA5.5), and each $\beta_d$ is the local average treatment effect (LATE) of an exogenous supply shock to a substantively interpretable content margin; Online Appendix (OA15.3) reports a joint nine-endogenous robustness check that confirms our main findings of engagement and commitment split.

\noindent\textbf{Controls.} To isolate the causal effects of these variables, our analysis includes a parsimonious set of time-varying controls. Two variables capture publisher-side exposure events: subscription prompts capture paywall banner exposures, and reduced-access articles record encounters with the publisher's registration wall when the user is not signed in. To distinguish polarization from sentiment, such as negativity, we include the average sentiment of consumed content (measured by an off-the-shelf transformer-based sentiment classifier trained for our focal language). As proxies for article quality and editorial depth, we add the weekly average article length and proper noun density; longer articles with more named entities typically reflect in-depth, investigative reporting rather than short opinion pieces. We also include weekly logins, the number of times a user authenticates in a given week\footnote{A registered user is recorded with or without an active sign-in.}, as a proxy for active platform use and as a prerequisite for most subscription actions. Finally, we control for non-political content consumption in our subscription and churn models, classified into eleven editorial categories drawn directly from the publisher's URL structure (e.g., .../sports/...), so that estimated effects load on polarizing content rather than general reading volume.

Table~\ref{tab:summary_stats} presents the summary statistics for the key variables in our panel of 1{,}098{,}857 user-week observations. The data exhibit features common to digital media consumption: user activity is highly skewed, with the mean value for time on site (3{,}044 seconds) being far above the 25th percentile (200 seconds). Polarization consumption is also highly skewed: at least the bottom quartile of user-weeks records zero clicks on every polarizing dimension, while the right tail extends to several hundred clicks per week for the most active users.

\begin{table}[!htbp]
\centering
\singlespacing
\caption{Summary Statistics of Key Variables (User-Week Level)}
\label{tab:summary_stats}
\begingroup
\setlength{\tabcolsep}{4.5pt}
\renewcommand{\arraystretch}{0.95}
\footnotesize
\begin{tabular}{l r S[table-format=4.3] S[table-format=4.3] S[table-format=3.1] S[table-format=4.1] r}
\toprule
& \multicolumn{1}{c}{N} & \multicolumn{1}{c}{Mean} & \multicolumn{1}{c}{SD} & \multicolumn{1}{c}{P25} & \multicolumn{1}{c}{P75} & \multicolumn{1}{c}{Max} \\
\midrule
\multicolumn{7}{l}{\textbf{Panel A: Dependent Variables}} \\
\addlinespace
Time on Site (Seconds) & 1{,}098{,}857 & 3044.259 & 6992.299 & 200 & 2994 & 1{,}093{,}988 \\
Subscription (0/1) & 1{,}050{,}215 & 0.006 & 0.079 & 0 & 0 & 1 \\
Churn (0/1) & 595{,}108 & 0.003 & 0.051 & 0 & 0 & 1 \\
\midrule
\multicolumn{7}{l}{\textbf{Panel B: Polarization Treatment Variables}} \\
\addlinespace
Affective Clicks (LLM) & 1{,}098{,}857 & 1.243 & 2.955 & 0 & 1 & 321 \\
Economic Left Clicks (LLM) & 1{,}098{,}857 & 1.088 & 2.397 & 0 & 1 & 200 \\
Economic Right Clicks (LLM) & 1{,}098{,}857 & 1.601 & 3.388 & 0 & 2 & 266 \\
Globalist Clicks (LLM) & 1{,}098{,}857 & 2.037 & 4.585 & 0 & 2 & 415 \\
Nationalist Clicks (LLM) & 1{,}098{,}857 & 1.388 & 3.280 & 0 & 2 & 304 \\
Pro-Environment Clicks (LLM) & 1{,}098{,}857 & 0.734 & 1.805 & 0 & 1 & 192 \\
Pro-Growth Clicks (LLM) & 1{,}098{,}857 & 0.287 & 0.925 & 0 & 0 & 147 \\
Left Clicks (BERT) & 1{,}098{,}857 & 2.684 & 5.297 & 0 & 3 & 401 \\
Right Clicks (BERT) & 1{,}098{,}857 & 2.582 & 5.106 & 0 & 3 & 365 \\
\midrule
\multicolumn{7}{l}{\textbf{Panel C: Control Variables}} \\
\addlinespace
Non-Political Content Clicks & 1{,}098{,}857 & 3.546 & 6.831 & 0 & 4 & 570 \\
Avg. Content Sentiment & 1{,}098{,}857 & {$-$0.128} & 0.200 & {$-$0.181} & {$-$0.003} & 1 \\
Avg. Article Length & 1{,}098{,}857 & 269.468 & 287.358 & 109.0 & 425.7 & 4848 \\
Avg. Proper Noun Density & 1{,}098{,}857 & 0.061 & 0.033 & 0.045 & 0.079 & 1 \\
Subscription Prompts & 1{,}098{,}857 & 0.117 & 0.792 & 0 & 0 & 87 \\
Logins & 1{,}098{,}857 & 0.275 & 1.621 & 0 & 0 & 220 \\
Reduced-Access Articles & 1{,}098{,}857 & 0.478 & 1.838 & 0 & 0 & 259 \\
\bottomrule
\end{tabular}
\vspace{0.3ex}
\begin{minipage}{\textwidth}
\scriptsize
\textit{Notes:} Time on Site is the sum of seconds spent on the platform across all sessions in a week; the raw weekly maximum of 1,093,988 seconds reflects multi-device, multi-tab, and idle-session artifacts in the tracking system, which the asinh transformation used in regressions compresses. The smaller $N$ for Subscription and Churn reflects their hazard-sample restrictions: Subscription is computed on user-weeks before the user's first paid order, and Churn on active-subscriber-weeks before any observed churn event. Variable definitions are detailed in Section~\ref{dependent-independent-variables}.
\end{minipage}
\endgroup
\end{table}
\doublespacing

\section{Empirical Strategy}\label{empirical-strategy}

This section develops our empirical strategy. We first specify a baseline fixed-effects model, then introduce two instrumental variables to address the endogeneity of content consumption, and finally validate both instruments through a series of diagnostic tests. 

To understand how polarizing content affects news consumption and subscription behavior, we begin by specifying the following fixed-effects ordinary least squares (OLS) regression as a baseline model:

\begin{equation}
\label{eqn:01-dim}
Y_{it} \;= \beta_{d}\, \mathrm{PolarizingArticles}_{itd}
\;+\; \alpha_{i} \;+\; \delta_{t} \;+\; \theta' X_{it} \;+\; \varepsilon_{it},
\end{equation}

In this model, \(Y_{it}\) is user \(i\)'s outcome in week \(t\), representing either user engagement (time on site, transformed with the inverse hyperbolic sine) or a linear-probability indicator of subscription and churn. \({PolarizingArticles}_{itd}\) is our main regressor, defined as the absolute number of clicks by user \(i\) in week \(t\) on articles classified along polarization dimension \(d\) constructed from the BERT (left or right) and LLM (affective; economic left/right; globalization globalist/nationalist; environmental pro-eco/pro-growth). \(\alpha_{i}\) are user fixed effects (capturing time-invariant heterogeneity such as baseline ideology or news taste),\footnote{We do not include article-author fixed effects because our identification operates at the user-week level, not the article level. User fixed effects absorb time-invariant reader-author matching (e.g., a reader who consistently follows a particular columnist), and the Bartik instrument's topic-week supply shifts aggregate across all articles in a topic-week cell, so no single author's production schedule drives the instrument.} and \(\delta_{t}\) are week fixed effects (capturing platform-wide shocks and seasonality). \(X_{it}\) is the vector of time-varying controls described above (sentiment, article quality proxies, subscription prompts, logins, reduced-access articles). We aggregate these events at the user-week level.

\subsection{Instrumental Variable (IV) Regression}\label{instrumental-variable-iv-regression}

Although Equation~\eqref{eqn:01-dim} accounts for user-fixed effects, time-fixed effects, and other time-varying characteristics, an OLS estimate of Equation~\eqref{eqn:01-dim} is unlikely to produce a causal estimate of the impact of polarizing articles because users choose to click on polarizing articles rather than being randomly assigned to do so.

To address this endogeneity problem, we utilize two separate and theoretically independent sources of exogenous variation: a demand-side shock from a major federal election and a supply-side Bartik-style instrument.

Our first instrument, the Bartik IV, is a supply-side, Bartik-style (or ``shift-share'') instrument \citep{bartik1991benefits,goldsmith2020bartik,bhj_2025_jep}. Originally developed to isolate exogenous variation in local labor demand, shift-share designs have been applied across a wide range of empirical settings, including platform markets \citep{CullenFarronato2021} and housing market effects of home-sharing \citep{BarronKungProserpio2021}. We define our Bartik IV as:

\begin{equation}
\label{eqn:03}
BartikIV_{itd} = \sum_k Share_{ik} \times Shift_{ktd},
\end{equation}
where $Share_{ik}$ is the share of topic $k$ in user $i$’s reading history during the pre-period (weeks 2--13), held fixed across all subsequent weeks, and $Shift_{ktd}$ is the week-$t$ supply shock for polarization dimension $d$ within topic $k$. In our primary (share-based) implementation, $Shift_{ktd}$ is measured from the publisher’s full archive as the \emph{share} of newly published articles in topic $k$ that are classified as polarizing along dimension $d$ that week, i.e., the topic-week share of polarizing content. Note that the endogenous treatment, $\mathrm{PolarizingArticles}_{itd}$, remains a count (clicks by user $i$ in week $t$ on articles scored as polarizing along dimension $d$); only the instrument's shift component is expressed in share units. We then interact these fixed shares with the weekly, topic-level ``shift''. Identification therefore comes from comparing weeks in which the publisher’s editorial mix within a topic shifts toward versus away from polarizing content, holding each user’s topic preferences and overall reading volume fixed; the resulting variation in exposure is driven by editorial choices and news shocks that are exogenous to any individual user's preferences. 

The Bartik instrument rests on two components, predetermined user shares and publisher-week topic shocks, each of which must exhibit sufficient variation. The cross-sectional distribution of users' pre-period topic preferences reveals substantial heterogeneity across content categories, and the weekly supply of polarizing articles displays pronounced time-series and cross-topic variation, generating wide dispersion of $\text{BartikIV}_{itd}$ across user-weeks. Additional diagnostics supporting the instrument's structure are reported in Online Appendix~OA6, including the effective number of shifts ($N_{\text{eff}} = 8.85$, indicating roughly nine independent shift sources across eleven publisher-desk content categories) and \citet{goldsmith2020bartik} projection $R^{2}$ values ranging from 0.94 to 0.99.
        Because $Share_{ik}$ is fixed in the clean pre-election window and $Shift_{ktd}$ varies at the publisher-week-topic level, the design targets relevance by construction and, conditional on week fixed effects, plausibly meets the exclusion restriction. 

Our second instrument, the Election IV, is a demand-side instrument that leverages a federal election that took place in our focal country (the publisher's ``home'' country) during the middle of our observation window (Week 20). Given the high reputation of our publisher in its linguistic region, our sample includes users who live outside the publisher's home country: 24\% of our registered users live abroad, among them, 83\% residing in a neighboring country that shares the same language but was not part of the election. These neighboring-country users serve as a credible control group, while users in our home country are in the treatment group. The instrument exploits differential demand rather than differential supply: as established in Section~\ref{publisher-setting}, the publisher does not geo-target content presentation, so any differential reading during weeks 20--21 reflects demand-side variation rather than platform-side curation. What separates the two groups is voting eligibility and exposure to parties' campaign communications, which raise home-country residents' demand for election coverage. This interaction between citizenship and election timing is plausibly exogenous to individual users' content preferences and to the publisher's editorial choices; the exclusion restriction is that being domestic during weeks 20--21 affects our engagement and monetization outcomes only through this induced shift in polarizing-content consumption. We therefore define 

\begin{equation}
\label{eqn:02}
ElectionIV_{it} = Country_{i} \times Election_{t},
\end{equation}
where $Country_{i} = 1$ if user $i$ is from our ``home'' country (0 otherwise), and $Election_{t} = 1$ during the 2-week federal election window (weeks 20--21), 0 otherwise. Thus, our $ElectionIV_{it}$ represents the interaction between the home country \({Country}_{i}\) and the election period \({Election}_{t}\). The \({Election}_{t}\) dummy is active during the election week (week 20) and the week immediately following (week 21). This window captures the peak of campaign salience and the immediate post-election period, when media coverage of the election campaign and its immediate aftermath is most intense\footnote{Polling day is the Sunday of week 20, so weeks 20--21 are the smallest two-week block containing the vote and its immediate aftermath. Online Appendix~OA9.5 confirms that the affective subscription and churn signs are stable across the alternative windows where the first stage is sufficiently strong.}. 


            We estimate the two instruments separately rather than in a single overidentified specification because each instrument identifies a distinct LATE on a different complier population. The Bartik IV captures the effect on users whose exposure responds to editorial supply across the 27-week estimation window (weeks 14--40); the Election IV captures the effect on users whose exposure responds to political salience during a concentrated two-week federal-election window (weeks 20--21). Because the complier populations, time horizons, and underlying mechanisms differ, the two LATEs need not coincide in magnitude, and need not fall on the same outcome margin: polarizing content can plausibly carry different commitment consequences in calm editorial weeks than during a high-salience election. 

            The first stage for each instrument is therefore defined as:
            \begin{equation}
            \mathrm{PolarizingArticles}_{itd} = \pi_{d}\ \cdot \text{Z}_{itd} + \alpha_{i} + \delta_{t} +\; \kappa' X_{it} \;+\; \nu_{it},
            \end{equation}
            where $Z_{itd}$ is either our $ElectionIV_{it}$ or our $BartikIV_{itd}$. Under the validity conditions we discuss next, two-stage least squares (2SLS) consistently estimates $\beta$, which captures the within-user, within-week effect of one additional polarizing click on our outcome measures. The second stage is given by:
            \begin{equation}
            Y_{it} = \beta_{d}\widehat{\mathrm{PolarizingArticles}}_{itd} + \alpha_{i} + \delta_{t} + \theta' X_{it} + \varepsilon_{it}.
            \end{equation}

\paragraph{Specification tiers.} We report two specification tiers. S1 includes no time-varying controls ($X_{it} = \emptyset$) and is our baseline specification, isolating the fixed-effects-only treatment effect. S5 adds all time-varying controls described above (sentiment, article length, proper noun density, subscription prompts, logins, and reduced-access articles), plus one additional control, non-political clicks, in subscription and churn models. We exclude non-political clicks from the engagement regression because the IV reallocates users across content categories, which makes this variable a potential post-treatment mediator: conditioning on it would block the substitution channel and induce bad-control bias. We include it in the subscription and churn models as a proxy for concurrent platform activity, while acknowledging that the same mediator concern could apply to those regressions too. Subscription prompts and reduced-access encounters raise similar concerns in the subscription specifications, since more polarizing reading could lift total volume and therefore the probability of hitting the metered threshold. To bound these concerns, we estimate both S1 and S5. As shown below, stability of the coefficient across S1 and S5 indicates that the core effect is neither largely confounded by concurrent article characteristics (like length or sentiment) nor primarily driven by downstream pathways (like general platform activity or paywall hits); intermediate specifications S2 -- S4 are reported in Online Appendix~OA12.3.

\paragraph{Estimation samples.} Two estimation samples are used throughout. For the Election IV, we use the \emph{full panel} (1{,}098{,}857 user-week observations across weeks 2--40; engagement regressions retain 1{,}070{,}735 after singleton drops by the user-and-week fixed-effects estimator), with standard errors clustered by user and week. For the Bartik IV, we restrict to users observed during the pre-period (weeks 2--13), whose topic shares are thereby predetermined, with estimation conducted on weeks 14--40 only \footnote{The 12-week pre-period ends well before the federal election (week 20), so user shares are fixed before any election-induced demand shift; the choice also balances share stability against estimation-window length. Online Appendix~OA10.1 reports robustness to alternative pre-period windows; All main findings are preserved.}. This restriction ensures that the shares entering the Bartik instrument are fixed before the estimation window, following the \citet{borusyak2022quasi} framework. The resulting estimation sample contains 544{,}246 user-week observations (541{,}588 after singleton drops), with standard errors clustered by user. Full descriptive statistics for this sample are reported in Online Appendix~OA14.

\subsection{Instrument Validity}\label{instrument-validity}

We assess instrument validity through a battery of diagnostics, summarized in Table~\ref{tab:iv_scorecard}. The OA column of the table gives the Online Appendix subsection where each test is reported in full; sections OA6--OA11 collect dimension-level results, implementation details, and supporting evidence.

\paragraph{Instrument strength.}
For the Bartik IV, the Montiel Olea--Pflueger effective first-stage $F$ \citep{montiel2013effective} ranges from 71 to 366 across all nine polarization dimensions and three outcomes, far above conventional weak-IV thresholds (rule-of-thumb $F=10$ of \citet{staiger1997instrumental}); the instrument draws on $N_{\text{eff}} = 8.85$ effective topic shocks (the demand-share Herfindahl--Hirschman index across 11 publisher-desk categories, following \citealp{borusyak2022quasi}, hereafter BHJ; see \hyperref[tab:hhi]{Table~OA6.1} in OA6.3).
For the Election IV, $F = 21.2$ on engagement and $18.6$ on subscription, both above conventional thresholds; the first stage on affective churn is weaker ($F = 7.0$, below the conventional $F = 10$ threshold), and we accordingly report Anderson--Rubin (AR) inference \citep{anderson1949estimation}. Full first-stage F-statistics for every dimension, outcome, and instrument are reported in Table~3 (S1 and S5) and Table~4 (S5); the complete S1--S5 ladder appears in OA12.3.

\paragraph{Exclusion restriction.}
For the Bartik IV, the principal diagnostic is the \citet{borusyak2022quasi} shock-balance test, which checks whether topic-week supply shocks are correlated with observable demand-side covariates that could drive outcomes through non-content channels. The test passes after Benjamini--Hochberg (BH) correction within each of the three outcome variables across the nine polarization dimensions (smallest $q \approx 0.34$; OA8.1). Three independent placebos converge on the same conclusion: weekly logins as a placebo outcome (0/9 dimensions reject; OA8.2), zero-click weeks as a non-reader placebo (8/9 null at $p > 0.10$; OA8.5), and reverse-causality regressions confirming that lagged demand does not predict supply (OA8.3). The leave-one-topic-out (LOTO) analysis (OA7.3) further shows that no single topic drives the identifying variation.

To build intuition for how the Bartik instrument could fail, consider a topic-specific threat: a financial crisis increases the publisher's supply of polarizing economics coverage (the supply shift) while simultaneously driving users with high pre-period economics reading shares to visit the platform more from heightened economic concern, independent of content tone. Week fixed effects absorb any aggregate demand shock common to all users, but the residual concern is that users whose pre-period shares load on the affected topic experience a differential engagement increase through non-content channels, biasing the engagement effect upward. Three of the diagnostics above are designed to detect exactly this failure mode. The BHJ shock-balance test (OA~8.1) attacks the precondition directly: it verifies that topic-week supply shocks are uncorrelated with observable demand-side covariates that could drive outcomes through non-content channels. The leave-one-topic-out analysis (OA~7.3) rules out any single event-driven topic dominating the identifying variation. The non-reader placebo (OA~8.5) verifies that users who read no polarizing content show no spurious response to the supply shifts, ruling out a demand-side response operating outside the content channel.

For the Election IV, the exclusion restriction holds if the election affects outcomes only through polarizing content consumption. Because the election date was set by national federal law years in advance, the timing of the shock is exogenous to platform behavior and to publisher decisions. Platform-wide shocks are absorbed by week fixed effects; time-invariant user heterogeneity by user fixed effects. A potential concern is that the election reduces subscription propensity among focal-country users through channels other than polarizing-content consumption: for example, if abundant free election coverage from broadcast media, social media, or party websites reduces the perceived value of a paid news subscription. Three features of our setting weigh against this alternative.

First, the subscription event study (OA~9.1) shows no differential pre-period decline between focal-country and control-country users, as would be expected if campaign coverage were already displacing subscription demand through non-content channels. Second, the engagement event study (OA~9.1) shows that focal-country platform engagement rises rather than falls around the election, the opposite of what a broad demand-displacement mechanism would predict if free election coverage elsewhere were drawing users away from the platform. Third, the sign of any residual bias from heightened civic interest is theoretically ambiguous: engaged citizens might value high-quality coverage more highly (raising willingness to pay), or free election coverage from broadcast and social media might saturate political-information demand and lower it. The S1-to-S5 near-invariance of the affective subscription coefficient ($\beta = -0.016$ to $-0.015$) is informative on this alternative channel. These S5 controls (logins, non-political reading volume, and paywall encounters) themselves respond to the election, so they are not strictly pre-treatment. Conditioning on them nevertheless leaves the coefficient unchanged, ruling out a general volume surge, topic reallocation, or a civic-interest channel mediated by overall engagement.

\paragraph{Pre-trends.}
For the Bartik IV, lead-placebo tests pass for eight of nine dimensions after Benjamini--Hochberg correction (the single failure is economic-right $\times$ engagement, $q < 0.001$). The rejection reflects topic-supply autocorrelation (lag-1 $\rho = 0.42$--$0.85$ across dimensions, mean 0.67), which is expected under persistent news cycles and does not invalidate the shift-share design \citep{borusyak2022quasi}. 

For the Election IV, engagement exhibits a known pre-trend reflecting systematically higher baseline engagement among domestic users beginning roughly one month before the election, as voting materials are progressively distributed to eligible residents. We address this pre-trend by including a group-specific quadratic time trend, and all 21 post-election weekly dummies survive the adjustment (OA~9.4). Subscription and churn pre-trends are clean (joint Wald $\chi^{2}$ tests in OA~9.1).

\paragraph{Robustness.}

The diagnostic battery in Table~\ref{tab:iv_scorecard} spans 19 tests across five categories: instrument strength (effective $F$, AR CIs), exclusion restriction (BHJ shock-balance, Conley $\gamma^{*}$, two placebos, parallel pre-trends, quadratic trend), exogeneity (reverse causality, projection $R^{2}$ following \citealp{goldsmith2020bartik}, hereafter GPSS), specification stability (LOTO, Rotemberg weights, pre-period window variation, election-window variation, progressive controls, two-way clustering), and additional shift-share diagnostics (BHJ shock-level 2SLS, cross-topic correlations, lead-placebo tests), with each row linking to the OA subsection where the test is reported in full. Beyond Table~\ref{tab:iv_scorecard}, OA15 reports four additional robustness exercises: a lagged-$Z$ test of the BHJ pre-determination assumption (OA15.1), randomization inference and wild-cluster bootstrap as design-free complements to clustered analytic SEs (OA15.2), a joint nine-endogenous 2SLS that adjusts each dimension's effect for cross-dimensional consumption co-occurrence (OA15.3), and \citet{rambachan2023more} pre-trend sensitivity for the Election engagement margin (OA15.4). Our main results are stable across this full battery of robustness tests.

\begin{table}[!htbp]
\centering
\singlespacing
\caption{Instrument Validation Summary}
\label{tab:iv_scorecard}
\footnotesize
\setlength{\tabcolsep}{5pt}
\renewcommand{\arraystretch}{0.92}
\begin{tabular}{p{4.0cm} p{7.2cm} c c c}
\toprule
\textbf{Test} & \textbf{What a pass confirms} & \textbf{Bartik} & \textbf{Election} & \textbf{OA} \\
\midrule
\multicolumn{5}{l}{\textit{Relevance: does the instrument move the endogenous variable?}} \\
Effective F-statistic     & Instruments are strong enough that IV is not biased toward OLS                  & \checkmark & \checkmark              & 7.1 \\
Anderson--Rubin CIs         & Conclusions hold under weak-instrument asymptotics                              & \checkmark & \checkmark              & 7.1 \\
\addlinespace[2pt]
\multicolumn{5}{l}{\textit{Exclusion: does the instrument affect outcomes only through the endogenous variable?}} \\
Shock-balance test (BHJ)    & Supply shocks are uncorrelated with unobserved demand shifters                  & \checkmark & n.a.                    & 8.1 \\
Conley sensitivity bounds   & Results survive even if the exclusion restriction is partially violated         & \checkmark & n.a.                    & 7.4 \\
Placebo outcomes (logins)   & Instrument does not predict outcomes it should not affect                       & \checkmark & n.a.                    & 8.2 \\
Non-reader placebo          & Instrument has no effect on users with zero political content exposure          & \checkmark & n.a.                    & 8.5 \\
Parallel pre-trends         & Subscription and churn pre-trends are clean; engagement pre-trend documented and absorbed via group-specific quadratic trend & n.a.       & \checkmark\textsuperscript{\dag}              & 9.1, 9.4 \\
Quadratic time trend        & Results are not driven by differential trends mistaken for the election         & n.a.       & \checkmark              & 9.4 \\
\addlinespace[2pt]
\multicolumn{5}{l}{\textit{Exogeneity: is the instrument as-good-as-random?}} \\
Reverse causality test      & Past demand does not drive current supply shifts                                & \checkmark & n.a.                    & 8.3 \\
GPSS projection $R^{2}$     & Pre-period shares capture actual topic preferences, not noise                   & \checkmark & n.a.                    & 6.4 \\
\addlinespace[2pt]
\multicolumn{5}{l}{\textit{Stability: are results robust to specification choices?}} \\
Leave-one-topic-out         & No single topic drives the entire result                                        & \checkmark & n.a.                    & 7.3 \\
Rotemberg weights           & No single topic overturns the result; negative weights are common in shift-share and do not invalidate the BHJ design & \checkmark & n.a.                    & 7.2 \\
Pre-period window variation & Main findings hold across alternative pre-period windows used to construct user shares & \checkmark & n.a.                    & 10.1 \\
Election window variation   & Commitment effects (subscription and churn) are sign-stable across alternative window definitions & n.a.       & \checkmark              & 9.5 \\
Progressive specification   & Results are stable as controls are added sequentially                           & \checkmark & \checkmark              & 12.3, 9.2 \\
Two-way clustering          & Standard errors remain adequate under cross-time error correlation              & \checkmark & \checkmark              & 7.5, 9.3 \\
\addlinespace[2pt]
\multicolumn{5}{l}{\textit{Additional robustness}} \\
BHJ shock-level 2SLS        & Micro-level and shock-level estimates are consistent                            & \checkmark & n.a.                    & 11.1 \\
Cross-topic correlations    & Supply shocks across topics are sufficiently independent                        & \checkmark & n.a.                    & 11.2 \\
Lead-placebo tests          & 8/9 dimensions pass after BH; one rejection reflects topic-supply autocorrelation under BHJ & \checkmark\textsuperscript{\dag} & n.a.                    & 8.4 \\
\bottomrule
\end{tabular}
\vspace{0.3ex}
\begin{minipage}{\textwidth}
\scriptsize
\textit{Notes:} \checkmark\ = test passes; \checkmark\textsuperscript{\dag}\ = pass with a documented qualification (one BH-flagged exception out of nine, or a single borderline cell), spelled out in the linked OA subsection; n.a.\ = not applicable (the test is designed for one instrument type and does not apply to the other; e.g., parallel pre-trends are not defined in a shift-share design, and shock-balance tests are not defined for an event-timing instrument). Conley $\gamma^{*}$ = \citet{conley2012plausibly}.
\end{minipage}
\end{table}

\section{Results}\label{results}

We present our results in three parts. Section~\ref{main-effects} reports the causal effects of affective polarization, Section~\ref{ideological-effects} extends the analysis to ideological polarization across multiple dimensions, and Section~\ref{mechanism} investigates the mechanism through engagement decomposition, the conversion funnel, and a test of whether reader demand follows confirmation bias or balanced consumption as the publisher's supply shifts.

\subsection{Main Effects of Affective Polarization}\label{main-effects}

\begin{table}[!htbp]
\centering
\singlespacing
\caption{The Causal Effect of Affective Polarization on User Behavior}
\label{tab:main_results_affective}
\resizebox{0.92\textwidth}{!}{%
\setlength{\tabcolsep}{4pt}
\begin{tabular}{l D{.}{.}{4.5} D{.}{.}{4.5} D{.}{.}{4.5} D{.}{.}{4.5} D{.}{.}{4.5} D{.}{.}{4.5}}
\toprule
 & \multicolumn{2}{c}{Engagement} & \multicolumn{2}{c}{Subscription} & \multicolumn{2}{c}{Churn} \\
\cmidrule(lr){2-3} \cmidrule(lr){4-5} \cmidrule(lr){6-7}
Dependent Var.: & \multicolumn{2}{c}{asinh(Time on Site)} & \multicolumn{2}{c}{Subscribed (0/1)} & \multicolumn{2}{c}{Churned (0/1)} \\
Instrument: & \multicolumn{1}{c}{Bartik IV} & \multicolumn{1}{c}{Election IV} & \multicolumn{1}{c}{Bartik IV} & \multicolumn{1}{c}{Election IV} & \multicolumn{1}{c}{Bartik IV} & \multicolumn{1}{c}{Election IV} \\
 & \multicolumn{1}{c}{(1)} & \multicolumn{1}{c}{(2)} & \multicolumn{1}{c}{(3)} & \multicolumn{1}{c}{(4)} & \multicolumn{1}{c}{(5)} & \multicolumn{1}{c}{(6)} \\
\midrule
\multicolumn{7}{l}{\textit{Specification S1 (no additional controls)}} \\[2pt]
Affective Content & 0.116^{*}   & 0.219       & 0.002       & -0.016^{**} & -0.001      & 0.036^{\dagger}  \\
                  & (0.059)     & (0.184)     & (0.001)     & (0.006)     & (0.001)     & (0.024)                 \\
                  &             &             &             &             &             & \multicolumn{1}{c}{\footnotesize$[0.005,\, 0.188]$} \\[4pt]
\multicolumn{7}{l}{\textit{Specification S5 (full controls)}} \\[2pt]
Affective Content & 0.116^{*}   & 0.060       & 0.002       & -0.015^{*} & -0.001      & 0.027^{\dagger}  \\
                  & (0.059)     & (0.242)     & (0.001)     & (0.006)     & (0.001)     & (0.020)                 \\
                  &             &             &             &             &             & \multicolumn{1}{c}{\footnotesize$[0.004,\, 0.217]$} \\
\midrule
First-Stage $F$ (S1)      & 255.8       & 21.2        & 232.6       & 18.6        & 203.8       & \multicolumn{1}{c}{7.0$^{\dagger}$}         \\
First-Stage $F$ (S5)      & 257.2       & 32.1        & 232.4       & 36.6        & 209.0       & \multicolumn{1}{c}{5.7$^{\dagger}$}         \\
Observations              & 541{,}588   & 1{,}070{,}735 & 513{,}646   & 1{,}019{,}140 & 336{,}714   & 583{,}856   \\[4pt]
User \& Week Fixed Effects & \multicolumn{1}{c}{Yes} & \multicolumn{1}{c}{Yes} & \multicolumn{1}{c}{Yes} & \multicolumn{1}{c}{Yes} & \multicolumn{1}{c}{Yes} & \multicolumn{1}{c}{Yes} \\
\bottomrule
\end{tabular}%
}
\vspace{0.3ex}
\begin{minipage}{\textwidth}
\scriptsize
\textit{Notes:} Each column reports a separate just-identified 2SLS regression of the outcome on the weekly count of clicks on articles scored as highly affective by the LLM scorer. S1 and S5 specifications, sample composition, hazard restrictions, and IV definitions follow Sections~\ref{dependent-independent-variables} and~\ref{empirical-strategy}. First-stage $F$ values are the \citet{montiel2013effective} effective F-statistic, computed under the same cluster-robust variance as the standard errors (equal to the cluster-robust Kleibergen--Paap rk Wald F \citep{kleibergen2006generalized} in this just-identified setting); this is the conservative cluster-robust diagnostic, roughly half the non-clustered alternative for the Election IV (OA7.1). $^{\dagger}$ marks $F<10$ (point estimate shown without significance stars; Anderson--Rubin 95\% CI in brackets); cells with $F<3.84$ are suppressed (``--'') because the AR confidence set is unbounded \citep{dufour1997some}. Cluster-robust standard errors for S5 binary outcomes apply a standard positive-semidefinite projection of the variance matrix; S1 specifications are unaffected. Standard errors clustered by user (Bartik) and two-way by user and week (Election). $^{*}\,p<0.05$; $^{**}\,p<0.01$; $^{***}\,p<0.001$.
\end{minipage}
\end{table}

We begin with affective polarization, for which our instruments proved most powerful. Table~\ref{tab:main_results_affective} reports the causal effects of consuming affectively polarizing content (LLM score $\geq$ 4) on engagement, subscription conversion, and churn. All models include user and week fixed effects; we report both S1 (no additional controls) and S5 (full control variables) to demonstrate robustness.

The results reveal an asymmetric trade-off. The Bartik IV identifies the engagement margin: affective polarization raises engagement ($\beta = 0.116$, $p = 0.049$, semi-elasticity $\approx 11.6\%$). Subscription, by contrast, does not move. With strong first stages ($F = 204$--$257$) and tightly estimated coefficients, the null is not an artifact of weak instruments but a genuine absence of subscription effects for supply-driven compliers. And the effect is not hiding in delayed response: a distributed-lag specification with two weekly lags yields a cumulative subscription effect statistically indistinguishable from zero (OA13.4.2). Even when the engagement boost is real and robust, that additional attention does not convert to subscriptions: under normal editorial operations, the engagement gains from polarizing content yield no detectable subscription revenue.

The Election IV identifies the commitment margin: affective polarization reduces subscription conversion ($\beta = -0.016$, $p = 0.008$) and accelerates churn (AR CI $[0.005,\, 0.188]$, $p = 0.017$). The same content attribute that raises engagement under supply-driven editorial variation reduces commitment under high political salience. Combined with the Bartik subscription null, the two instruments together show that polarizing content at best wastes the subscription funnel and at worst erodes it.

The two IVs identify LATEs on different complier populations: Bartik IV on supply-driven compliers, where engagement rises but commitment is unaffected; Election on salience-driven compliers, where the commitment penalty emerges sharply (engagement point estimate positive but imprecise, $\beta = 0.219$, $p = 0.24$). Together, they rule out a scenario in which polarizing content benefits both margins.

\subsection{Main Effects of Ideological Polarization}\label{ideological-effects}

Table~\ref{tab:main_results_ideological} presents the causal effects of consuming ideologically polarizing content on engagement, subscription conversion, and churn, estimated separately for each dimension using both instruments. All estimates use the S5 specification (full controls); S1 results (reported in full in OA12) are qualitatively similar.
Panel~A presents BERT-classified ideological polarization. The Bartik IV reveals an engagement asymmetry: right-leaning content raises engagement ($\beta = 0.081$, $p = 0.03$), while left-leaning content reduces it ($\beta = -0.137$, $p = 0.04$) and raises churn ($\beta = +0.004$, $p = 0.036$); readers are drawn toward right-leaning over left-leaning content. The Election IV shows that right-leaning content during the high-salience period also reduces subscription ($\beta = -0.009$, $p = 0.0014$) and accelerates churn (AR CI $[0.005,\, \infty)$, $p = 0.018$, $F = 5.1$; with a weak first stage, the AR CI establishes the sign of the churn effect but does not bound its magnitude). Across the two instruments, BERT-classified content imposes a commitment penalty in both ideological directions: the right via the engagement--commitment trade-off (engagement up under Bartik, commitment down under Election), the left via lower engagement and higher churn under Bartik.
\begin{table}[!htbp]
\centering
\singlespacing
\caption{The Causal Effect of Ideological Polarization on Engagement, Subscription, and Churn}
\label{tab:main_results_ideological}
\renewcommand{\arraystretch}{0.82}
\setlength{\tabcolsep}{4pt}
\resizebox{0.92\textwidth}{!}{%
\begin{tabular}{l D{.}{.}{4.5} D{.}{.}{4.5} D{.}{.}{4.5} D{.}{.}{4.5} D{.}{.}{4.5} D{.}{.}{4.5}}
\toprule
 & \multicolumn{2}{c}{Engagement} & \multicolumn{2}{c}{Subscription} & \multicolumn{2}{c}{Churn} \\
\cmidrule(lr){2-3} \cmidrule(lr){4-5} \cmidrule(lr){6-7}
Dependent Var.: & \multicolumn{2}{c}{asinh(Time on Site)} & \multicolumn{2}{c}{Subscribed (0/1)} & \multicolumn{2}{c}{Churned (0/1)} \\
Instrument: & \multicolumn{1}{c}{Bartik} & \multicolumn{1}{c}{Election} & \multicolumn{1}{c}{Bartik} & \multicolumn{1}{c}{Election} & \multicolumn{1}{c}{Bartik} & \multicolumn{1}{c}{Election} \\
 & \multicolumn{1}{c}{(1)} & \multicolumn{1}{c}{(2)} & \multicolumn{1}{c}{(3)} & \multicolumn{1}{c}{(4)} & \multicolumn{1}{c}{(5)} & \multicolumn{1}{c}{(6)} \\
\midrule
\multicolumn{7}{l}{\textbf{Panel A. BERT-Based Ideological Polarization}} \\[1pt]
Effect of Left Clicks & -0.137^{*} & \multicolumn{1}{c}{--} & 0.001 & \multicolumn{1}{c}{--} & 0.004^{*} & \multicolumn{1}{c}{--} \\
 & (0.065) & & (0.002) & & (0.002) & \\
\addlinespace[0.15em]
Effect of Right Clicks & 0.081^{*} & 0.039 & 0.000 & -0.009^{**} & -0.000 & 0.033^{\dagger} \\
 & (0.037) & (0.160) & (0.001) & (0.003) & (0.001) & (0.022) \\
 & & & & & & \multicolumn{1}{c}{\footnotesize$[0.005,\, \infty)$} \\
\midrule
\multicolumn{7}{l}{\textbf{Panel B. LLM-Based Ideological Polarization}} \\[1pt]
Effect of Econ-Left & 0.008 & \multicolumn{1}{c}{--} & 0.002 & \multicolumn{1}{c}{--} & -0.003 & \multicolumn{1}{c}{--} \\
 & (0.063) & & (0.001) & & (0.002) & \\
\addlinespace[0.15em]
Effect of Econ-Right & 0.029 & 0.029^{\dagger} & 0.000 & -0.008^{\dagger} & -0.001 & 0.015^{\dagger} \\
 & (0.050) & (0.129) & (0.001) & (0.002) & (0.001) & (0.006) \\
 & & \multicolumn{1}{c}{\footnotesize$[-0.136,\, \infty)$} & & \multicolumn{1}{c}{\footnotesize$[-\infty,\, -0.005]$} & & \multicolumn{1}{c}{\footnotesize$[0.004,\, 0.036]$} \\
\addlinespace[0.15em]
Effect of Globalist & -0.068 & \multicolumn{1}{c}{--} & 0.000 & \multicolumn{1}{c}{--} & 0.001 & \multicolumn{1}{c}{--} \\
 & (0.051) & & (0.001) & & (0.001) & \\
\addlinespace[0.15em]
Effect of Nationalist & 0.001 & 0.052^{\dagger} & 0.000 & -0.013^{**} & 0.000 & 0.021^{*} \\
 & (0.035) & (0.216) & (0.001) & (0.004) & (0.001) & (0.010) \\
 & & \multicolumn{1}{c}{\footnotesize$[-0.367,\, 0.768]$} & & & & \\
\addlinespace[0.15em]
Effect of Enviro-Left & 0.082 & \multicolumn{1}{c}{--} & -0.001 & \multicolumn{1}{c}{--} & 0.001 & 0.064^{\dagger} \\
 & (0.051) & & (0.001) & & (0.001) & (0.039) \\
 & & & & & & \multicolumn{1}{c}{\footnotesize$[0.010,\, 0.323]$} \\
\addlinespace[0.15em]
Effect of Enviro-Right & 0.199 & 0.206^{\dagger} & -0.001 & -0.053^{***} & 0.001 & \multicolumn{1}{c}{--}  \\
 & (0.132) & (0.885) & (0.002) & (0.014) & (0.003) &  \\
 & & \multicolumn{1}{c}{\footnotesize$[-1.050,\, 4.610]$} & & & & \\
\midrule
\multicolumn{7}{l}{\textit{Diagnostics: First-Stage F-statistic (Montiel Olea--Pflueger Effective F)}} \\
Panel A: Left / Right & \multicolumn{1}{c}{111 / 291} & \multicolumn{1}{c}{0.18$^{\dagger}$ / 16.06} & \multicolumn{1}{c}{71 / 247} & \multicolumn{1}{c}{0.84$^{\dagger}$ / 25.68} & \multicolumn{1}{c}{80 / 220} & \multicolumn{1}{c}{0.21$^{\dagger}$ / 5.08$^{\dagger}$} \\
Panel B: Econ L / R & \multicolumn{1}{c}{295 / 366} & \multicolumn{1}{c}{0.03$^{\dagger}$ / 3.90$^{\dagger}$} & \multicolumn{1}{c}{311 / 311} & \multicolumn{1}{c}{0.06$^{\dagger}$ / 4.58$^{\dagger}$} & \multicolumn{1}{c}{204 / 230} & \multicolumn{1}{c}{0.83$^{\dagger}$ / 8.30$^{\dagger}$} \\
Panel B: Glob L / R & \multicolumn{1}{c}{197 / 273} & \multicolumn{1}{c}{1.94$^{\dagger}$ / 9.19$^{\dagger}$} & \multicolumn{1}{c}{255 / 212} & \multicolumn{1}{c}{0.51$^{\dagger}$ / 10.73} & \multicolumn{1}{c}{126 / 184} & \multicolumn{1}{c}{1.56$^{\dagger}$ / 19.40} \\
Panel B: Env L / R & \multicolumn{1}{c}{256 / 91} & \multicolumn{1}{c}{0.50$^{\dagger}$ / 8.36$^{\dagger}$} & \multicolumn{1}{c}{259 / 83} & \multicolumn{1}{c}{0.72$^{\dagger}$ / 10.65} & \multicolumn{1}{c}{313 / 152} & \multicolumn{1}{c}{6.54$^{\dagger}$ / 0.001$^{\dagger}$} \\
\midrule
Observations (Bartik / Election) & \multicolumn{1}{c}{541{,}588} & \multicolumn{1}{c}{1{,}070{,}735} & \multicolumn{1}{c}{513{,}646} & \multicolumn{1}{c}{1{,}019{,}140} & \multicolumn{1}{c}{336{,}714} & \multicolumn{1}{c}{583{,}856} \\
User \& Week FE & \multicolumn{1}{c}{Yes} & \multicolumn{1}{c}{Yes} & \multicolumn{1}{c}{Yes} & \multicolumn{1}{c}{Yes} & \multicolumn{1}{c}{Yes} & \multicolumn{1}{c}{Yes} \\
Controls Included & \multicolumn{1}{c}{Yes} & \multicolumn{1}{c}{Yes} & \multicolumn{1}{c}{Yes} & \multicolumn{1}{c}{Yes} & \multicolumn{1}{c}{Yes} & \multicolumn{1}{c}{Yes} \\
\bottomrule
\end{tabular}%
}
\vspace{0.3ex}
\begin{minipage}{\textwidth}
\scriptsize
\textit{Notes:} Each row reports a separate just-identified 2SLS regression on the weekly click count for the indicated polarization dimension under S5 (full controls). Sample composition, hazard filtering, and clustering follow Table~\ref{tab:main_results_affective}. Second-stage estimates are suppressed (--) when first-stage $F<3.84$, the $\chi^{2}(1)$ critical value below which the Anderson--Rubin confidence set is unbounded \citep{dufour1997some}; for $3.84\le F<10$ ($^{\dagger}$), the point estimate is shown without significance stars and the AR 95\% confidence interval is reported in brackets. $^{*}\,p<0.05$; $^{**}\,p<0.01$; $^{***}\,p<0.001$.
\end{minipage}
\end{table}

Panel~B shows results for the six LLM ideological dimensions. The Election IV's first-stage power, where it exists, concentrates on right-leaning content; left-leaning F values fall uniformly below $F = 10$ (eight of nine below $F = 2$). This mirrors the campaign itself: the federal election was won by a right-wing populist party whose nationalist and anti-immigration agenda dominated coverage, concentrating the election-induced demand shift on right-leaning supply. Engagement effects are null across all six dimensions under S5; the engagement margin is carried by affective and BERT-right content rather than by these finer-grained ideological cuts. The commitment margin, however, is broader. During the election period, nationalist content reduces subscription ($\beta = -0.013$, $p = 0.0015$) and accelerates churn ($\beta = 0.021$, $p = 0.044$); environmental-right content reduces subscription ($\beta = -0.053$, $p < 0.001$); and economic-right content accelerates churn (AR CI $[0.004,\, 0.036]$, $p = 0.018$, $F = 8.3$). Wherever the Election IV has first-stage power, polarizing content lowers willingness to pay or raises willingness to leave, indicating that the commitment penalty propagates across ideological dimensions even where engagement does not measurably move.

Taken together, the engagement margin of polarizing content is narrow, with only affective and BERT-classified right-leaning content drawing readers in, while the commitment margin is broader, propagating under the Election IV across affective and right-leaning ideological content during the high-salience period. The directional pattern is consistent across our two measurement frameworks: under both BERT and LLM, right-leaning content carries the Election-IV commitment penalty. Under Benjamini–Hochberg correction across the eight ideological dimensions at S5, the four right-leaning ideological dimensions surviving for Election subscription at $q \le 0.005$ come from both pipelines (BERT-Right from BERT; Nationalist, Economic-Right, and Environmental-Right from LLM); engagement and churn do not survive multiple-testing correction at this specification (\hyperref[tab:oa12_4_bh]{Table~OA12.8}). This is a robustness that no single instrument or polarization measure could deliver. Polarization is selectively engaging but broadly costly to subscriber commitment.

\subsection{Mechanism}\label{mechanism}

Our main results establish a divergence: affective polarization raises engagement under the Bartik IV but does not convert to subscriptions; during the high-salience election window, it actively reduces subscription conversion and accelerates churn. To understand why the additional attention does not translate into commitment, we examine three mechanism channels: whether affective exposure deepens per-article reading or merely broadens low-depth browsing, whether the extra engagement progresses through the publisher's monetization funnel, and whether these effects reflect confirmation bias or the balanced-consumption alternative documented in the literature. Table~\ref{tab:mechanism_results} organizes the evidence into four panels: Panel~A decomposes the engagement margin into views, time per article, and active days; Panel~B traces the conversion funnel through weekly subscription prompts and an article-level paywall regression-discontinuity design; Panel~C tests confirmation bias from the user side using three pre-determined ideology proxies as moderators; and Panel~D tests confirmation bias and balanced consumption from the supply side using within-dimension counter-side Bartik shocks. The panel regressions use specification S1, the baseline column of Table~\ref{tab:main_results_affective} that anchors the headline engagement effect ($\beta = 0.116$). We use S1 rather than the S5 specification because several of the S5 controls (logins, reduced-access encounters, non-political clicks) are themselves potential mediators, so conditioning on them would absorb part of the channels we decompose. Full diagnostics of mechanisms are in Online Appendix~OA13.
\begin{table}[!htbp]
    \centering
    \begingroup
    \singlespacing
    \caption{Mechanism Evidence: Engagement Margins, Conversion Path, Formal Moderators, and Direct Ideological Supply}
    \label{tab:mechanism_results}
    \renewcommand{\arraystretch}{0.85}
    \footnotesize
    \setlength{\tabcolsep}{6pt}
    {%
        \begin{tabular}{l D{.}{.}{4.7} D{.}{.}{4.7}}
            \toprule
            \multicolumn{3}{l}{\textbf{Panel A: Engagement margins (affective polarization, 2SLS)}} \\
            \cmidrule(l){1-3}
            Outcome (asinh)          & \multicolumn{1}{c}{Bartik IV} & \multicolumn{1}{c}{Election IV} \\
            \midrule
            Total time on site       & 0.116^{*}            & 0.219                \\
                                     & (0.059)              & (0.184)              \\
            Total article views      & 0.049^{**}           & 0.727^{***}          \\
                                     & (0.019)              & (0.114)              \\
            Time per article         & 0.065                & -0.446^{**}          \\
                                     & (0.049)              & (0.142)              \\
            Active days              & -0.005               & 0.026                \\
                                     & (0.011)              & (0.065)              \\
            \midrule
            \multicolumn{3}{l}{\textbf{Panel B: Weekly funnel and article-level paywall threshold}} \\
            \cmidrule(l){1-3}
            Outcome                  & \multicolumn{1}{c}{Weekly Bartik IV} & \multicolumn{1}{c}{Threshold RD} \\
            \midrule
            subscription-prompt incidence                        & -0.024^{**}  & \multicolumn{1}{c}{--} \\
                                                        & (0.0076)     &                         \\
            Prompt jump at article~11                   & \multicolumn{1}{c}{--} & 0.00493^{***}  \\
                                                        &                         & (0.00047)      \\
            Purchase jump at article~11                 & \multicolumn{1}{c}{--} & 0.000315^{**}  \\
                                                        &                         & (0.000106)     \\
            \midrule
            \multicolumn{3}{l}{\textbf{Panel C: User-side ideology moderators (interaction term only)}} \\
            \cmidrule(l){1-3}
            Moderator (affective $\times$)                            & \multicolumn{1}{c}{Bartik IV} & \multicolumn{1}{c}{Election IV} \\
            \midrule
            Residential right-leaning zip (indicator), engagement     & 0.069        & -0.434       \\
                                                                      & (0.039)      & (0.616)      \\
            Residential right-leaning zip (indicator), subscription   & -0.0006      & -0.0069      \\
                                                                      & (0.0009)     & (0.0124)     \\
            Pre-period reading lean (LLM), engagement                 & -0.133       & -0.729       \\
                                                                      & (0.086)      & (1.344)      \\
            Pre-period reading lean (LLM), subscription               & -0.0019      & -0.0069      \\
                                                                      & (0.0018)     & (0.0106)     \\
            Pre-period reading lean (BERT), engagement                & -0.099       & -0.112       \\
                                                                      & (0.060)      & (0.891)      \\
            Pre-period reading lean (BERT), subscription              & 0.0020       & 0.031        \\
                                                                      & (0.0013)     & (0.044)      \\
            \midrule
            \multicolumn{3}{l}{\textbf{Panel D: Counter-side reading share (within-dimension Bartik reduced form)}} \\
            \cmidrule(l){1-3}
            Dimension                & \multicolumn{2}{c}{Bartik reduced form}        \\
            \midrule
            Economic                 & \multicolumn{2}{c}{$1.578^{***}$ \quad (0.056)} \\
            Globalization            & \multicolumn{2}{c}{$1.203^{***}$ \quad (0.058)} \\
            Environmental            & \multicolumn{2}{c}{$2.638^{***}$ \quad (0.101)} \\
            \bottomrule
        \end{tabular}%
    }
    \vspace{0.3ex}
    \begin{minipage}{\textwidth}
        \scriptsize
        \textit{Notes:} Panel~B threshold-RD: bandwidth $h=5$ articles, linear local slope ($p=1$); $h$ denotes the half-width of the RD window in articles, $\tau$ the RD treatment effect at article~11. Sample composition and hazard filtering follow Table~\ref{tab:main_results_affective}; standard errors clustered by user (Bartik and RD rows) and two-way by user and week (Election rows). Panel~C reports the interaction coefficient $\beta_2$ from the just-identified 2SLS specification $Y_{it} = \beta_1 A_{it} + \beta_2 (A_{it} \times M_i) + \alpha_i + \delta_t + \varepsilon_{it}$, where $A_{it}$ is the affective treatment and $M_i$ is the time-invariant ideology moderator (residential zip vote share, weeks~2--13 LLM lean, or weeks~2--13 BERT lean); both $A_{it}$ and $A_{it} \times M_i$ are instrumented with the affective Bartik (or Election) instrument and its interaction with $M_i$. Residential battery: every affective $\times$ residential-zip interaction has $q > 0.6$ under BH adjustment ($q$ = Benjamini--Hochberg adjusted $p$-value across the multiple-testing battery); pre-period-lean battery: every interaction $|t| < 2$, raw. Panel~D reports the reduced-form pass-through coefficient $\gamma_d$ on the counter-side Bartik shock from $\text{counter-share}_{itd} = \gamma_d Z^{\text{counter}}_{itd} + \zeta_d Z^{\text{own}}_{itd} + \alpha_i + \delta_t + \varepsilon_{itd}$, where $Z^{\text{counter}}_{itd}$ is the Bartik shock to publisher-side supply of content opposite the user's pre-period lean on axis $d$ and $Z^{\text{own}}_{itd}$ is the analogous own-side supply shock entered as a control. $^{*}\,p<0.05$; $^{**}\,p<0.01$; $^{***}\,p<0.001$.

    \end{minipage}
    \endgroup
\end{table}

Panel~A decomposes the engagement response. Affective exposure raises total article views under both instruments (Bartik $\beta = 0.049$, $p = 0.010$; Election $\beta = 0.727$, $p < 0.001$), but per-article depth does not follow: time per article is indistinguishable from zero under Bartik and significantly negative under Election ($\beta = -0.446$, $p = 0.002$), and active days are null in both designs. Put behaviorally, readers drawn to affective content open more articles per session but do not spend more time inside each article and do not return to the site on more distinct days; the engagement margin is a wider per-session shelf-reach, not a deeper or more habitual relationship. The same pattern appears for BERT-classified right-leaning ideological content under Election ($\beta_{\text{views}} = 0.450$, $\beta_{\text{time/art}} = -0.276$; OA13.1.1 Panel~B), broadening the engagement-trap pattern beyond the affective dimension to right-leaning ideological supply during the high-salience period. The mix of news categories also shifts, but in opposite directions across the two designs (OA13.1.2): Bartik compliers reshuffle toward business and technology reading, while Election compliers concentrate on domestic political content. Individual articles consumed under Bartik are also shorter, more negative, and more emotional (OA13.1.3).

Panel~B asks whether this additional engagement pushes users deeper into the commitment funnel. Two complementary designs converge in rejecting this mechanism. In the weekly Bartik panel, restricted to the weeks over which weekly subscription-prompt incidence is tracked (weeks 23 through 40), affective exposure reduces the probability of receiving a subscription prompt ($\beta = -0.024$, $p = 0.002$). That is, polarizing content yields fewer encounters with the paywall rather than more. To isolate the causal effect of the prompt at the gate itself, we exploit the publisher's two-tier metered paywall: a lower registration boundary after the fifth free article for unregistered readers (registered readers see ten free articles) and the article-11 subscription wall, where the 11th article in a user-month triggers a subscription prompt for registered readers. Readers who reach exactly 10 articles in a month and readers who reach 11 are very similar, so any sharp jump in subscription behavior between them isolates the causal effect of the prompt rather than of differences in reading intensity. This is a regression discontinuity (RD) design. We compare outcomes within a bandwidth of $h = 5$ articles on either side of the subscription threshold (articles 6 through 16), fitting a line on each side and absorbing user-month fixed effects so that each user serves as her own comparison within her own month. The preferred $h=5$ RD window covers article counts 6 through 16; its lower endpoint sits at, not below, the registration boundary, so the preferred estimate is not contaminated by the article-5 placebo gradient documented in OA13.2.4. Bandwidth sensitivity to $h = 3$ (narrower, similar point estimate) and $h = 8$ (which crosses the registration wall and accordingly yields a larger estimate) is reported in OA13.2.2. The sample is restricted to the post-election four-month window in which article-level prompt tracking is fully operational (359{,}897 user-months and roughly 1.85 million article visits). The probability of receiving a prompt jumps by $\tau = 0.00493$ at the article-11 threshold ($t = 10.56$), and purchase probability rises about 10\% over a 0.33\% baseline ($\tau = 0.000315$, $t = 2.98$). Critically, no LLM or BERT polarization measure moderates the purchase jump in any specification we test (OA13.2.3): across nine polarization dimensions and the full battery of interaction, content-composition, and split-sample tests, no result survives multiple-testing correction. The conversion response at the threshold is invariant to article polarization. Because the meter-crossing decision is itself a behavioral selection, these estimates describe a self-selected, high-intent population rather than a strict causal effect of the prompt alone; the manipulation diagnostic and full RD robustness checks are in OA13.2.4.

Why does affective content that captures attention fail to convert readers into subscribers? Three channels are consistent with our evidence, though we cannot definitively distinguish among them. First, \emph{perceived quality erosion}: us--versus--them rhetoric may signal editorial opinion rather than the investigative depth that readers associate with a newspaper of record, reducing the perceived value of a paid subscription. Consistent with this, Bartik-shifted affective consumption coincides with shorter, more negative, and more emotional articles (OA13.1.3), a content mix that resonates with recent evidence that negative and emotionally charged news draws disproportionate consumption \citep{RobertsonEtAl2023NHB}, yet the opposite profile of the long-form analytical content that typically motivates subscription. Second, \emph{perceived bias}: encountering confrontational framing may lead readers to infer that the publisher is advancing an agenda rather than providing balanced coverage, undermining the impartiality that subscribers expect from a newspaper of record, independent of the content's depth or analytical quality. This channel is distinct from perceived quality erosion because it concerns readers' inferences about editorial position, not the depth or informational value of any single article: a subscription is not payment for a single engaging story but a continuing relationship with a trusted information institution. Third, \emph{engagement without attachment}: affective content may function as low-commitment ``snacking'' that satisfies curiosity without deepening the reader's relationship with the news platform, consistent with the Panel~A finding that article views rise but time per article and active days do not. All three channels predict the pattern we observe (engagement up, commitment down), and all three are consistent with the trap being conditional on political salience, when affective content is most abundant and most psychologically salient.

Panels~C and D return to the puzzle posed in the introduction: does demand follow confirmation bias or balanced consumption? The two panels identify the same mechanism from opposite sides of the news market. Panel~C asks whether a reader's prior ideology amplifies the affective response (demand-side test); Panel~D asks whether an exogenous shift in counter-side supply passes through to counter-side reading (supply-side test). Under confirmation bias, Panel~C should show a non-zero moderator and Panel~D's pass-through should be near zero; under balanced consumption, the predictions reverse for both panels. Panel~C interacts the affective treatment with three pre-determined ideology proxies: residential zip-code vote shares (exogenous to reading by construction), and weeks~2--13 reading lean classified independently by our LLM and our BERT pipelines, so that classifier-specific measurement error in one framework cannot drive a spurious moderator estimate that survives in the other. The interaction and the affective treatment are instrumented jointly in a just-identified 2SLS specification (see Table~\ref{tab:mechanism_results} notes). Across a full battery of moderator tests, spanning engagement, subscription, and churn under both instruments (OA13.3.1 for geographic, OA13.3.2 for pre-period reading lean), no ideology interaction systematically moderates the affective effect; ideologically congruent and incongruent users move together. This null, however, has asymmetric precision: the largest Election-IV interaction standard error (1.344, on LLM lean $\times$ Election engagement) is over an order of magnitude larger than the main affective effect, while the Bartik-IV interactions are estimated much more tightly. We therefore read these moderator nulls as evidence against ideologically heterogeneous responses comparable to or larger than the main affective effect, while the design remains statistically uninformative about smaller heterogeneity. On the affective dimension, where the Election design has first-stage power, a separate test of reading composition runs against confirmation bias: left-leaning readers \emph{increase} rather than retreat from incongruent reading ($\beta = +0.094$, $p < 0.001$; OA13.3.3).

Panel~D tests the same hypothesis from the supply side. For each user, on each of three ideological dimensions where the publisher carries both sides (Economic, Globalization, Environment), we classify the user's weeks~2--13 reading as Left- or Right-leaning on that axis and define counter-side reading as the within-axis share of articles opposite that lean. We then estimate the reduced-form pass-through of an exogenous Bartik shock to counter-side supply onto the user's counter-side reading share, controlling for the analogous own-side supply shock (specification in Table~\ref{tab:mechanism_results} notes). Confirmation bias predicts a near-zero pass-through coefficient: readers screen out incongruent content even when more of it is supplied. Balanced consumption predicts a positive pass-through: readers absorb the additional counter-side supply. We find positive, precisely estimated pass-throughs on all three dimensions: counter-side reading share rises by $+1.578$ on Economic, $+1.203$ on Globalization, and $+2.638$ on Environmental per unit of the supply shock (each $|t| > 19$; OA13.3.4). Because each dimension uses its own Left- vs. Right-leaning classification and its own Bartik instrument, measurement error on any single dimension cannot drive this pattern; the data reject the confirmation-bias prediction in favor of balanced consumption. This response concentrates among readers whose pre-period reading was already balanced rather than one-sided, and it does not crowd out own-side reading, so within-axis total consumption expands rather than substitutes (OA13.3.5). Together, Panels~C and D address the demand-side puzzle: the engagement boost and the subscription penalty are not driven by readers preferentially consuming what reinforces their priors; where supply can be cleanly instrumented, readers rebalance toward the counter side rather than self-reinforce.

\section{Conclusion}\label{summary-and-conclusion} 
This study estimates causal effects of polarizing content on user engagement (time on site) and commitment (subscriptions and retention), the two revenue-relevant behaviors for digital news publishers. Our analysis combines detailed demand and supply data from a major European publisher with a novel BERT- and LLM-based framework for measuring polarization and two complementary instruments. We highlight three sets of findings and then discuss their implications for publishers, regulators, and platforms.

Our central finding is a ``polarization trap'' for affectively polarizing content: a Bartik-identified engagement gain under supply variation paired with an Election-identified commitment penalty under high salience, rather than a single unified effect on the same margin. The Bartik IV identifies an engagement gain alongside a precisely estimated null on commitment, so supply-driven engagement gains do not convert to subscriptions. The Election IV reveals that under high political salience the same content attribute reduces subscription and accelerates churn. Polarizing content captures attention, but we find no evidence that it helps the subscription funnel, and under high salience it hurts it. The trap is not specific to the affective dimension: BERT-classified right-leaning content also carries both margins, and during the election the commitment penalty extends to right-leaning ideological dimensions. The engagement margin is narrow; the commitment margin is broad. Mechanism analysis finds no statistically significant evidence of ideology-based heterogeneity in the engagement-commitment trade-off across three pre-determined proxies. The direct positive evidence comes from the supply side: on three ideological dimensions covering both sides, users rebalance toward content opposite their pre-period lean as the publisher's supply shifts, a pattern consistent with balanced consumption rather than confirmation bias.

The strategic dilemma these findings present for publishers can be made concrete with a simple back-of-the-envelope calculation. Our Election IV estimates indicate that an additional click on a highly affective article decreases the weekly subscription conversion probability by 1.6 percentage points (see Table~\ref{tab:main_results_affective}, Column~4). This effect is large relative to the weekly subscription base rate of 0.63\% (more than two and a half times the base rate), consistent with a Local Average Treatment Effect identified on compliers during a high-salience political window rather than an average effect across all users and weeks. The Election first-stage shift for the instrumented domestic election-window user-week is 0.150 additional affective clicks (Online Appendix~OA12), so the implied reduced-form subscription drop for that user-week is approximately 0.24 percentage points ($0.150 \times 0.016$). Assuming a conservative lifetime value of a new subscriber of \$420 (e.g., \$35/month for an average tenure of 12 months), the expected loss in subscription revenue from one additional polarizing click is approximately \$6.72 ($0.016 \times \$420$). The marginal ad revenue from one user's increased engagement in a single week (a few extra ad impressions at typical CPMs, worth roughly \$0.01--\$0.03) cannot offset this expected loss. Even at the most conservative end of the 95\% confidence interval (upper bound $= -0.0042$, computed from $\beta = -0.016$, SE $= 0.006$), the implied loss is \$1.76 per click, still roughly 59--176 times the marginal ad revenue. Even if the population-average effect is substantially smaller than the LATE, the implied revenue loss remains economically significant given the high lifetime value of each converted subscriber. The polarization trap is therefore not merely a statistical phenomenon but a first-order economic one.

A natural question is why polarizing content appears to be on the rise across news media despite these subscription costs. Our results suggest three complementary explanations. First, the polarization trap is difficult to detect with standard editorial analytics. Naive OLS estimates show uniformly positive associations between polarizing clicks and both engagement and subscription (Online Appendix~OA7.6); only after instrumenting for the endogeneity of content choice does the negative subscription effect emerge. An editor relying on observational dashboards would conclude, incorrectly, that polarizing content converts readers into subscribers. Second, the trade-off is business-model-specific. For outlets that remain primarily ad-funded, the engagement boost from polarizing content translates into revenue with no offsetting subscription cost; the finding bites hardest for publishers transitioning toward subscription-dependent models, including most major legacy outlets. Third, external distribution incentives compound the problem: social media algorithms and news aggregators reward high-engagement content, creating supply-side pressure that may overwhelm subscription-focused editorial strategies even when publishers are aware of the trade-off.

Our findings carry implications for three audiences. First, for subscription-focused publishers, our results suggest that a content strategy relying on affectively polarizing content is risky for subscription value. Such content captures fleeting attention without translating into subscription gains under everyday supply variation. During periods of high political salience, it appears to erode the trust and perceived value needed to convert readers into paying subscribers and to retain existing ones. Editorial teams that optimize on engagement metrics (clicks, time on site, sharing) may inadvertently undermine subscription conversion, the very revenue stream increasingly central to the financial sustainability of quality journalism. The actionable margin is not whether to cover politically salient events but how: the same event can be framed with varying degrees of affective intensity, and our findings indicate that editorially restraining us--versus--them rhetoric, even on inherently political topics, may protect subscription conversion without sacrificing topical relevance.

Second, our findings speak to active regulatory debates about digital content and consumer welfare. A core assumption behind engagement-based content optimization (on news sites, social media feeds, and recommendation algorithms alike) is that engagement reflects consumer preferences. Our results challenge this assumption: users click on and spend time with content that lowers their likelihood of subscribing and raises their likelihood of leaving, a pattern consistent with revealed-preference regret in which short-run consumption choices diverge from longer-run valuations. The underlying question raised here, whether engagement metrics adequately proxy for consumer welfare, generalizes beyond our setting and is directly relevant to debates about algorithmic amplification on digital platforms, though our setting is an editorially curated news publisher rather than a social media feed.

Third, our findings point to a distinct tension for digital platform operators. The issue is not only that engagement may be an incomplete proxy for consumer welfare, but that it may also be an incomplete objective function for platforms whose business depends on durable user relationships. On platforms where ranking and recommendation systems amplify content that maximizes clicks, comments, shares, or dwell time, affectively charged material, popularly known as ``rage bait,'' may raise measured engagement today while lowering willingness to pay, retention, creator support, or trust tomorrow. In that sense, engagement optimization can be privately self-defeating even before any welfare or regulatory rationale is invoked. Oxford University Press's selection of ``rage bait'' as its 2025 Word of the Year, alongside a reported threefold increase in usage over the preceding year \citep{oxford2025ragebait}, underscores the public salience of this attention strategy, even as its longer-run economic consequences remain understudied. The trade-off should be sharpest for platforms that monetize both attention and commitment: social media services with premium tiers, creator platforms with recurring subscriptions, paid communities, and membership-based aggregators. Testing whether the engagement-commitment divergence we document in an editorially curated news setting extends to algorithmically curated feeds is therefore a natural target for future research.

Two limitations merit discussion. First, our IV estimates should be interpreted as LATEs for the respective complier populations identified by each instrument, as discussed in Section~\ref{empirical-strategy}. Second, our data come from a single European newspaper of record whose audience spans the political spectrum, raising the question of whether the polarization trap operates similarly at outlets with ideologically self-selected audiences, for instance a right-wing or populist outlet. Our mechanism analysis (Section~\ref{mechanism}) offers a partial answer: the engagement-commitment trade-off does not vary with users' own ideology under three pre-determined proxies, and readers' consumption rises with exogenous shifts in the publisher's supply of content opposite their pre-period lean, consistent with balanced consumption rather than confirmation bias. Audiences self-selected into partisan outlets may nevertheless have higher tolerance for us--versus--them rhetoric, or affective intensity may erode trust even among sympathetic readers beyond some threshold; resolving this requires data from outlets spanning the ideological spectrum, including tabloid, populist, and broadcast media, as well as publishers operating in two-party systems. Two design features partially offset the single-publisher concern: two theoretically independent instruments identify complementary margins of a single trade-off (the Bartik IV the engagement gain on supply-shifted compliers, the Election IV the commitment penalty during a high-salience demand window), and the directional pattern holds across two distinct measurement frameworks (BERT and LLM) and multiple polarization dimensions, a robustness that a single instrument or single polarization measure could not deliver.

\begin{doublespace}
Future research could expand on our findings in several directions. First, exploring the interplay between content features, such as the interaction of headline tone with body content, could clarify the mechanisms of the polarization trap. Second, combining our demand-side estimates with data on advertising yields and content production costs could enable a structural analysis of the publisher's optimal polarization level, and, by extension, whether editorial or ideological objectives lead outlets to deviate from a profit-maximizing content mix. Third, our retention analysis could be extended with richer duration models that track how cumulative exposure to polarizing content shapes subscriber retention over longer horizons. Fourth, our single-publisher design cannot address competitive dynamics: if rival outlets simultaneously increase polarization, the subscription cost we estimate may be attenuated (readers have fewer balanced alternatives) or amplified (market-wide trust erosion in news media). Multi-publisher data would enable analysis of the equilibrium polarization level across a media market. Together, these extensions would sharpen what our findings already suggest: the economics of digital news media can incentivize content that, while engaging, undermines publishers' financial sustainability and perceived trustworthiness.
\end{doublespace}

\bigskip
\section*{References}
\addcontentsline{toc}{section}{References}
\begingroup
\singlespacing
\small
\setlength{\bibsep}{0.2ex plus 0.1ex}
\renewcommand{\bibsection}{}
\bibliographystyle{apalike}
\bibliography{bibtex}
\endgroup

\end{document}